\documentclass[12pt,twoside,british,english,prd,nofootinbib,preprint,showpacs,preprintnumbers,floatfix]{revtex4-1}
\usepackage{mathptmx}
\usepackage{helvet}
\usepackage{courier}
\usepackage[T1]{fontenc}
\usepackage[latin9]{inputenc}
\usepackage[a4paper]{geometry}
\usepackage{array}
\usepackage{float}
\usepackage{amsmath}
\usepackage{amssymb}
\usepackage{graphicx}
\usepackage{verbatim}
\usepackage{babel}

\makeatletter

\providecommand{\tabularnewline}{\\}
\newcommand{\lyxdot}{.}

\@ifundefined{textcolor}{}
{%
 \definecolor{BLACK}{gray}{0}
 \definecolor{WHITE}{gray}{1}
 \definecolor{RED}{rgb}{1,0,0}
 \definecolor{GREEN}{rgb}{0,1,0}
 \definecolor{BLUE}{rgb}{0,0,1}
 \definecolor{CYAN}{cmyk}{1,0,0,0}
 \definecolor{MAGENTA}{cmyk}{0,1,0,0}
 \definecolor{YELLOW}{cmyk}{0,0,1,0}
}

\makeatother

\begin{document}
\begin{flushright}
TTK-15-32 
\par\end{flushright}

\vspace{4mm}

\begin{center}
\textbf{\LARGE{}{}{}Prospects of constraining the Higgs CP nature
}{\LARGE{}{} } 
\par\end{center}

\begin{center}
\textbf{\LARGE{}{}{}in the tau decay channel at the LHC}{\LARGE{}{} } 
\par\end{center}

\begin{center}
\vspace{6mm}

\par\end{center}

\begin{center}
\textbf{\large{}{}{}Stefan Berge}{\large{}{}}%
\footnote{\texttt{\small{}{}{}berge@physik.rwth-aachen.de}%
}\textbf{\large{}{}{},}{\large{}{} }\textbf{\large{}{}{}Werner
Bernreuther}{\large{}{}}%
\footnote{\texttt{\small{}{}{}breuther@physik.rwth-aachen.de}%
}\textbf{\large{}{}{} and Sebastian Kirchner}{\large{}{}}%
\footnote{\texttt{\small{}{}{}kirchner@physik.rwth-aachen.de}%
}\textbf{\large{}{}{} }{\large{}{} } 
\par\end{center}

\begin{center}
Institut für Theoretische Teilchenphysik und Kosmologie, \\ RWTH Aachen
University, 52056 Aachen, Germany 
\par\end{center}

\begin{center}
\vspace{17mm}
 \textbf{Abstract} 
\par\end{center}

We investigate how precisely  the CP nature of the
 125 GeV Higgs boson $h$, parametrized by a
scalar-pseudoscalar Higgs mixing angle  $\phi_\tau$,
 can be determined in   $h\to \tau^-\tau^+$ decay
 with subsequent $\tau$-lepton decays to charged prongs
  at the Large Hadron Collider (LHC).
 We combine  two methods in order to define an observable $\varphi^*_{CP}$ 
 which is sensitive to  $\phi_\tau$: We use the $\rho$-decay plane method  for $\tau^\mp\to\rho^\mp$
 and the impact parameter method for all other major $\tau$ decays. 
 For estimating the precision with which  $\phi_\tau$ can be measured at the LHC (13~TeV) we take into account
 the  $\tau^-\tau^+$ background from Drell-Yan production and perform  a  Monte Carlo  simulation
 of  measurement uncertainties  on the   $\varphi^*_{CP}$ signal and background distributions.
 We obtain that  the mixing angle $\phi_\tau$ can be determined with an uncertainty of 
     $\Delta\phi_\tau\simeq 15^{\circ}$
 ($9^{\circ}$) at the LHC with an integrated luminosity of $150\, {\rm fb}^{-1}$ $(500\, {\rm fb}^{-1})$,
 and with $\Delta\phi_\tau\approx 4^{\circ}$ with  $3\, {\rm ab}^{-1}$.
 Future measurements of  $\phi_\tau$ yield direct information on whether or not there is
 an extended Higgs-boson sector with  Higgs-sector CP violation. We analyze this in the context
 of a number of two-Higgs-doublet extensions of the Standard Model, namely the so-called aligned model
 and conventional two-Higgs-doublet extensions with tree-level neutral flavor conservation. 

\vspace{20mm}

  \newpage{}

\section{Introduction}
\label{sec:intro}

The investigations of the production and decay modes of the $h(125{\rm GeV})$ spin-zero resonance, which was discovered at the Large Hadron Collider
 (LHC) in 2012~\cite{Aad:2012tfa,Chatrchyan:2012ufa}, show  that the properties of this particle
  are compatible \cite{Khachatryan:2014jba,Aad:2015gba} with those of the Standard Model (SM) Higgs boson. 
   In particular, the analysis of angular correlations in $h\to Z Z, W^+W^-$ excluded that
   $h$ is a pseudoscalar state $(J^P=0^-)$  \cite{Chatrchyan:2012jja,Aad:2013xqa}. However, the results
    of \cite{Chatrchyan:2012jja,Aad:2013xqa}
 do not imply that $h$ is purely CP-even $(J^P=0^+)$, because a pseudoscalar component of $h$ 
 is most likely not detectable in its decays to weak gauge bosons. 

 The exploration whether or not $h(125{\rm GeV})$ is a pure CP-even state is of prime interest. If a CP-odd component of $h$ would be detected, that is, if $h$
  would turn out to be a CP mixture, it would be evidence of new physics, i.e., of non-standard CP violation. 
  One way to probe the CP nature of $h$ is to measure $\tau\tau$ spin correlations in $h\to\tau^+\tau^-$. 
       Respective phenomenological investigations for $h$ production and decay to $\tau$ leptons at the LHC include
    \cite{Berge:2008wi,Berge:2008dr,Berge:2011ij,Harnik:2013aja,Przedzinski:2014pla,Dolan:2014upa,Berge:2014sra,Askew:2015mda}.
    Because the $\tau^+\tau^-$ zero-momentum frame (ZMF) and 
     the $\tau^\pm$ rest frames cannot be experimentally reconstructed at the LHC, the central aspect of these analyses is to 
      define an alternative inertial frame and to construct an observable in this frame which allows to discriminate with high 
       sensitivity between a CP-even, CP-odd, and a CP-mixed Higgs boson $h$.
     The method proposed and applied in \cite{Berge:2008dr,Berge:2011ij,Berge:2013jra,Berge:2014sra} is applicable to all subsequent $\tau^\pm$ decays into 
      1 or 3 charged prongs. It requires the measurement of the energy and 3-momentum of the charged prong (which in our case is either
       a charged lepton or a charged pion)
      and its impact parameter in the laboratory frame. We will call this approach the impact parameter method in the following. 
      Another method, which we will call the $\rho$-decay plane method, was first proposed and applied in~\cite{Bower:2002zx,Desch:2003mw,Desch:2003rw,Was:2002gv,Worek:2003zp}
       for Higgs-boson production and decay to $\tau^+\tau^-$ in $e^+e^-$ collisions. The method works  only for subsequent $\tau$ decays to
      charged $\rho$ mesons and requires the measurement of the 4-momenta of the charged and neutral pion from $\rho^\pm$ decay. This method was analyzed 
       in \cite{Harnik:2013aja,Przedzinski:2014pla,Askew:2015mda} for Higgs-boson production at the LHC and shows a better sensitivity to 
        scalar-pseudoscalar Higgs mixing than the  impact parameter method applied to $\tau\to \rho$ decays.  
        
   The aim of this paper is to combine both methods. We apply a slight variant of the $\rho$-decay plane method to $\tau^\pm\to \rho^\pm$ and 
     the impact parameter method of \cite{Berge:2008dr,Berge:2011ij,Berge:2013jra,Berge:2014sra} 
     to the other major 1- and 3-prong $\tau$ decays and define a discriminating variable for probing the CP nature 
      of $h$. We  analyze in this way all major $\tau$ decays into one and three charged prongs. We estimate the statistical uncertainty with which the
      scalar-pseudoscalar Higgs mixing angle $\phi_\tau$ defined below can eventually be measured at the LHC (13 TeV) with this approach.  
         Assuming that this  precision on  $\phi_\tau$  can be reached, we investigate its impact on the parameters of a number of 
          Standard Model   extensions with non-standard CP violation, in particular 
            Higgs-sector CP violation,  namely the so-called aligned two-Higgs-doublet model \cite{Pich:2009sp}
          and conventional two-Higgs-doublet models with  neutral flavor conservation but Higgs-sector CP violation at tree-level.

 In this paper we consider the production of  the Higgs boson $h(125{\rm GeV})$ at the LHC with 13~TeV center-of-mass energy
  and its subsequent decay into a pair of $\tau$ leptons:
  \begin{align}
 pp\to h+X\to\,\tau^{-}\tau^{+}+X & .
 \label{eq:h2tautau_decay}
 \end{align}
The analysis of this paper can be applied to any $h$ production mode. For definiteness we shall consider below 
 Higgs production by gluon gluon fusion. The general model-independent effective Yukawa interactions of $h$ with $\tau$ leptons 
  can be parametrized as follows:
\begin{equation}
{\cal L}_{Y}=-\frac{m_{\tau}}{\rm v}~\kappa_{\tau}\left(\cos\phi_{\tau}\bar{\tau}\tau+\sin\phi_{\tau}\bar{\tau}i\gamma_{5}\tau\right)h\,,
\label{YukLa-phi}
\end{equation}
where ${\rm v}=246$ GeV, $\kappa_{\tau}>0$ denotes the reduced Yukawa
coupling strength, and the angle $\phi_{\tau}$ parametrizes the amount
of CP violation in the $h\tau\tau$ interaction. Henceforth, we will call $\phi_{\tau}$ the scalar-pseudoscalar
 Higgs mixing angle.
 At this point we recall the following terminology. We call  
 a neutral Higgs boson to be a CP-even or scalar state  (CP-odd or pseudoscalar state) 
  if it couples -- also beyond the tree level -- only to scalar (pseudoscalar) fermion currents.
   If the Higgs boson couples to both currents we call it a CP mixture.

  As already mentioned above, we analyze all major 1- and 3-prong tau decay modes:
\begin{eqnarray}
\tau & \to & l+\nu_{l}+\nu_{\tau}\,,\label{eq:taul} \\
\tau & \to & a_{1}+\nu_{\tau}\to\pi+2\pi^{0}+\nu_{\tau}\,,  \label{eq:taua1} \\
\tau & \to & a_{1}^{L,T}+\nu_{\tau} \to2\pi^{\pm}+\pi^{\mp}+\nu_{\tau} \,, \label{eq:taua1LT}  \\
\tau & \to & \rho+\nu_{\tau}\to\pi+\pi^{0}+\nu_{\tau}\,, \label{eq:taurho}\\
\tau & \to & \pi+\nu_{\tau}\, . \label{eq:taupi}
\end{eqnarray}
We call the decay mode $\tau\to a_{1}^{L,T}+\nu_{\tau}$  in \eqref{eq:taua1LT} also `1-prong',
because the 4-momentum of $a_{1}^{\pm}$ can be obtained from the
measured 4-momenta of the 3 charged pions. The longitudinal~$(L)$
and transverse $(T)$ helicity states of the $a_{1}$ resonance can
be separated by using known kinematic distributions \cite{Rouge:1990kv,Davier:1992nw,Kuhn:1995nn,Stahl:2000aq}.

In the following we denote the final charged particles by $a^{-},a'^{+}\in\{e^{\pm},\mu^{\pm},\pi^{\pm},a_{1}^{L,T,\pm}\}$.
 The normalized distributions of polarized $\tau^\mp$ decays to $a^\mp$ are, in the
  $\tau^\mp$ rest frame, of the form:
 \begin{eqnarray}
{\Gamma_{a}}^{-1}\mbox{d}\Gamma_a\left(\tau^{\mp}(\hat{{\bf
      s}}^{\mp})\to a^{\mp}(q^{\mp})+X\right) & = &
n\left(E_{\mp}\right)\left[1\pm b\left(E_{\mp}\right)\,\hat{{\bf
      s}}^{\mp}\cdot\hat{{\bf
      q}}^{\mp}\right]dE_{\mp}\frac{d\Omega_{\mp}}{4\pi} \, . 
      \label{eq:dGamma_dEdOmega}
\end{eqnarray}
Here, ${\bf \hat{s}}^{\mp}$ are the normalized spin vectors of
the $\tau^{\mp}$ and $E\mp$ and $\hat{{\bf q}}^{\mp}$ are the energies
and unit 3-momenta of $a^{\mp}$ in the respective
$\tau$ rest frame. The spectral functions $n$ and $b$ are given, for instance,
in \cite{Berge:2011ij}. The function $b(E_{\mp})$ contains the information on the $\tau$-spin
analyzing power of the particle $a^{\mp}$. We recall that the $\tau$-spin analyzing
power is maximal for the direct decays to pions, $\tau^{\mp}\to\pi^{\mp}$,
and for $\tau^{\mp}\to a_{1}^{L,T,\mp}.$ (The $\tau$-spin
analyzing power of $a_{1}^{L-}$ and $a_{1}^{T-}$ is $+1$ and $-1$,
respectively.) For the other decays, the $\tau$-spin analyzing power
of $l^{\mp}$ and $\pi^{\mp}$ depends on the energy of these particles.
It can be enhanced by appropriately chosen energy cuts.

The paper is organized as follows. In Sec.~\ref{sec:observables}  we first review the impact parameter method
 of \cite{Berge:2008dr,Berge:2011ij,Berge:2013jra,Berge:2014sra}.
 Then we  introduce a slightly modified version of the $\rho$-decay plane  method~\cite{Bower:2002zx,Desch:2003mw,Desch:2003rw,Was:2002gv,Worek:2003zp}
  which allows to combine both methods for those $\tau^-\tau^+$ decays where one $\tau$ lepton decays to $\rho + \nu_\tau$ and the other
   one to a charged prong $a \neq \rho$. We define an angle $\varphi_{CP}^{*}$ with which one can probe, with this combined method
   and for all $\tau^-\tau^+$    decay modes listed above, 
      whether or not $h$ has a CP-violating coupling to the $\tau$ lepton. Moreover, we define an asymmetry \cite{Berge:2014sra,Berge:2013jra} that is useful in estimating
       the  error $\Delta\phi_\tau$ with which the mixing angle $\phi_\tau$ can be measured in 
      each  $\tau^-\tau^+$ decay channel.
       In  Sec.~\ref{sec:phiestimate} we apply the combined method introduced in the previous section to the 
       $h\to\tau\tau$ decay modes listed above, at the LHC (13 TeV). We take into account the irreducible background from
        Drell-Yan production, $\gamma^*/Z^*\to \tau^-\tau^+$, apply acceptance cuts and account for measurement uncertainties by Monte Carlo
         simulation\footnote{Our own Monte-Carlo simulation program uses the external 
         software packages \cite{vanHameren:2007pt,Vermaseren:2000nd,Buckley:2014ana,Gough:2009:GSL:1538674,Brun:1997pa}.}
         as in \cite{Berge:2014sra}. We estimate the precision with which  the mixing angle $\phi_\tau$ can eventually be 
          measured at the LHC (13 TeV).    In Sec.~\ref{sec:constrH} we investigate the impact 
           this precision on $\phi_\tau$ and the expected precision on the reduced Yukawa coupling strength 
           $\kappa_\tau$ would have on the parameters of Standard Model extensions with non-standard CP violation, in particular
            Higgs-sector CP violation.
           We confine ourselves to non-supersymmetric two-Higgs-doublet extensions. First we analyze the 
            so-called aligned two-Higgs-doublet model~\cite{Pich:2009sp} and then discuss conventional 
      two-Higgs-doublet models (2HDM)  with tree-level neutral flavor conservation and a CP-violating tree-level Higgs potential, namely 
      the type-I and type-II 2HDM, the flipped, and the lepton specific model. Finally, we add a short remark on the so-called inert model. 
      We conclude in Sec.~\ref{sec:conlusions}.

\section{Observables}
\label{sec:observables}
In the decay $h\to\tau\tau$ 
 the information on the scalar-pseudoscalar mixing angle $\phi_{\tau}$
is encoded in the spin-spin correlation of the $\tau^{+}\tau^{-}$
leptons. For $\beta_{\tau}=\sqrt{1-4m_{\tau}^{2}/m_{h}^{2}}\approx1$ 
the differential decay width is proportional to (cf., for instance~\cite{Berge:2011ij})
\begin{align}
d\Gamma_{h\to\tau^{+}\tau^{-}} & \propto~ 1-s_{z}^{-}s_{z}^{+}+\cos\left(2\phi_{\tau}\right)\left({\bf s}_{T}^{\,-}\cdot{\bf s}_{T}^{\,+}\right)\nonumber \\
 & \qquad\quad{}+\sin\left(2\phi_{\tau}\right)\left[\left({\bf s}_{T}^{\,+}\times{\bf s}_{T}^{\,-}\right)\cdot{\bf \hat{k}}^{-}\right]\,,
 \label{dGamma_s-s+}
\end{align}
where ${\bf \hat{k}}^{-}$ is the normalized $\tau^{-}$ 3-momentum
in the Higgs-boson rest frame, ${\bf \hat{s}}^{\mp}$ are the unit
spin vectors of the $\tau^{\mp}$ in their respective $\tau$ rest
frames\footnote{These $\tau$ rest frames are
obtained from the Higgs rest frame by a rotation-free Lorentz boost
along the ${\bf \tau}^{\pm}$ momenta.},
 and $s_{z}^{\mp}$ $({\bf s}_{T}^{\,\mp})$ denotes the longitudinal (transverse) 
  component of ${\bf \hat{s}}^{\mp}$
with respect to ${\bf \hat{k}}^{-}$.
Eq.~\eqref{dGamma_s-s+} shows that, 
in the Higgs-boson rest frame, information on $\phi_{\tau}$ is obtained from the
correlation of the transverse components of the $\tau$-spins. This correlation is encoded 
 in the distribution of the angle between the plane defined by ${\bf \hat{s}}^{-}$
and ${\bf \hat{k}}^{-}$ and the plane defined by ${\bf \hat{s}}^{+}$
and ${\bf \hat{k}}^{-}$. The $\tau$ leptons self-analyze their spin direction through their
 parity-violating weak decays into charged prongs (cf. \eqref{eq:dGamma_dEdOmega}).
 Nevertheless, the angle between the above-mentioned plane cannot be measured directly because
  the $\tau^\pm$ rest-frames cannot be reconstructed. 
 Yet, the impact parameter method \cite{Berge:2008dr,Berge:2011ij,Berge:2013jra,Berge:2014sra}
or the $\rho$-decay 
 plane method~\cite{Bower:2002zx,Desch:2003mw,Desch:2003rw,Was:2002gv,Worek:2003zp} allows to
  determine this angle without reconstruction of the  4-momenta of the $\tau^\mp$.
 
\subsection{Impact parameter method}
\label{susec:impactp}

\begin{figure}[tb]
\noindent \raggedright{}\hspace*{1cm}\includegraphics[height=6cm]{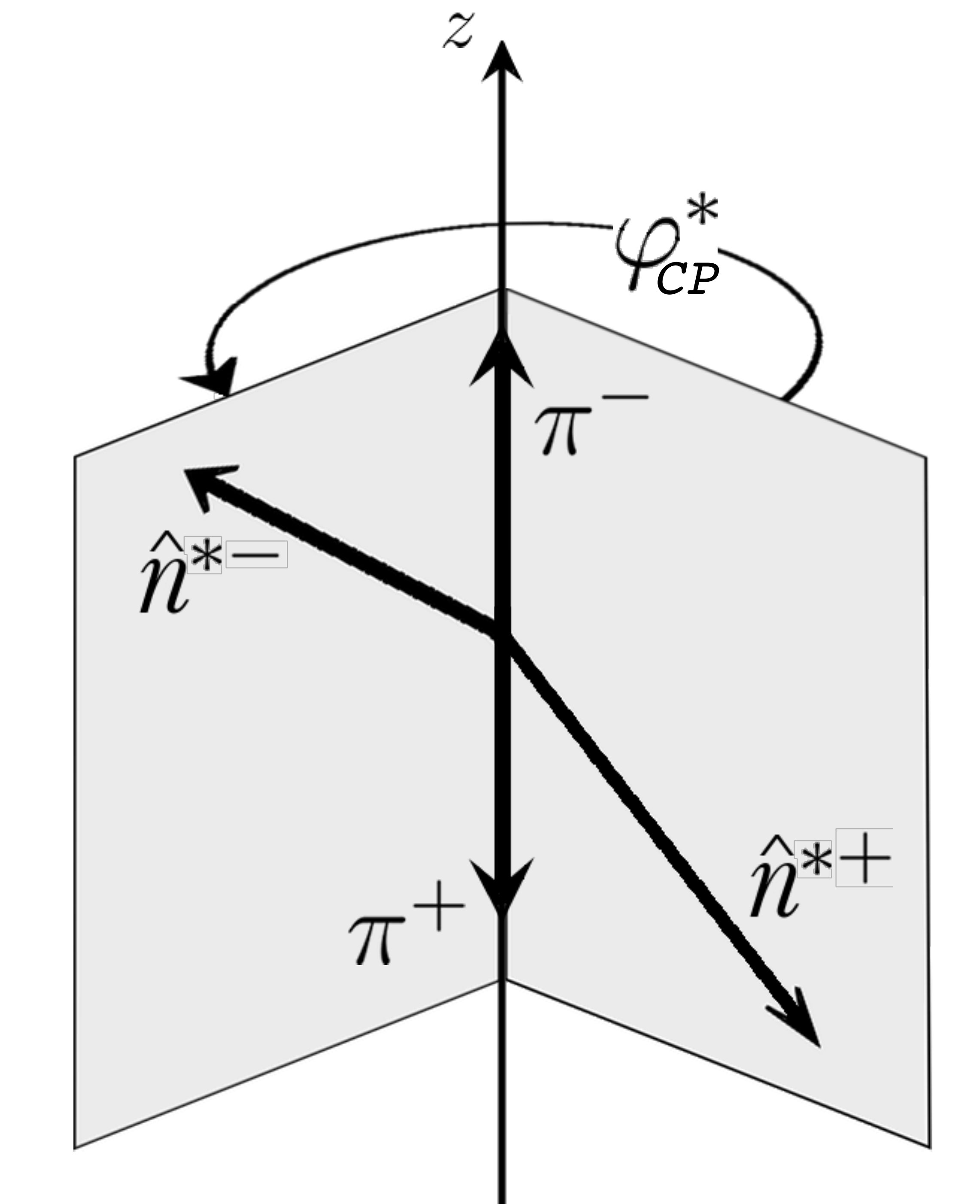}
\hspace*{0cm}\vspace{-5.3cm} \\
 \hspace*{6.7cm}\includegraphics[height=6cm]{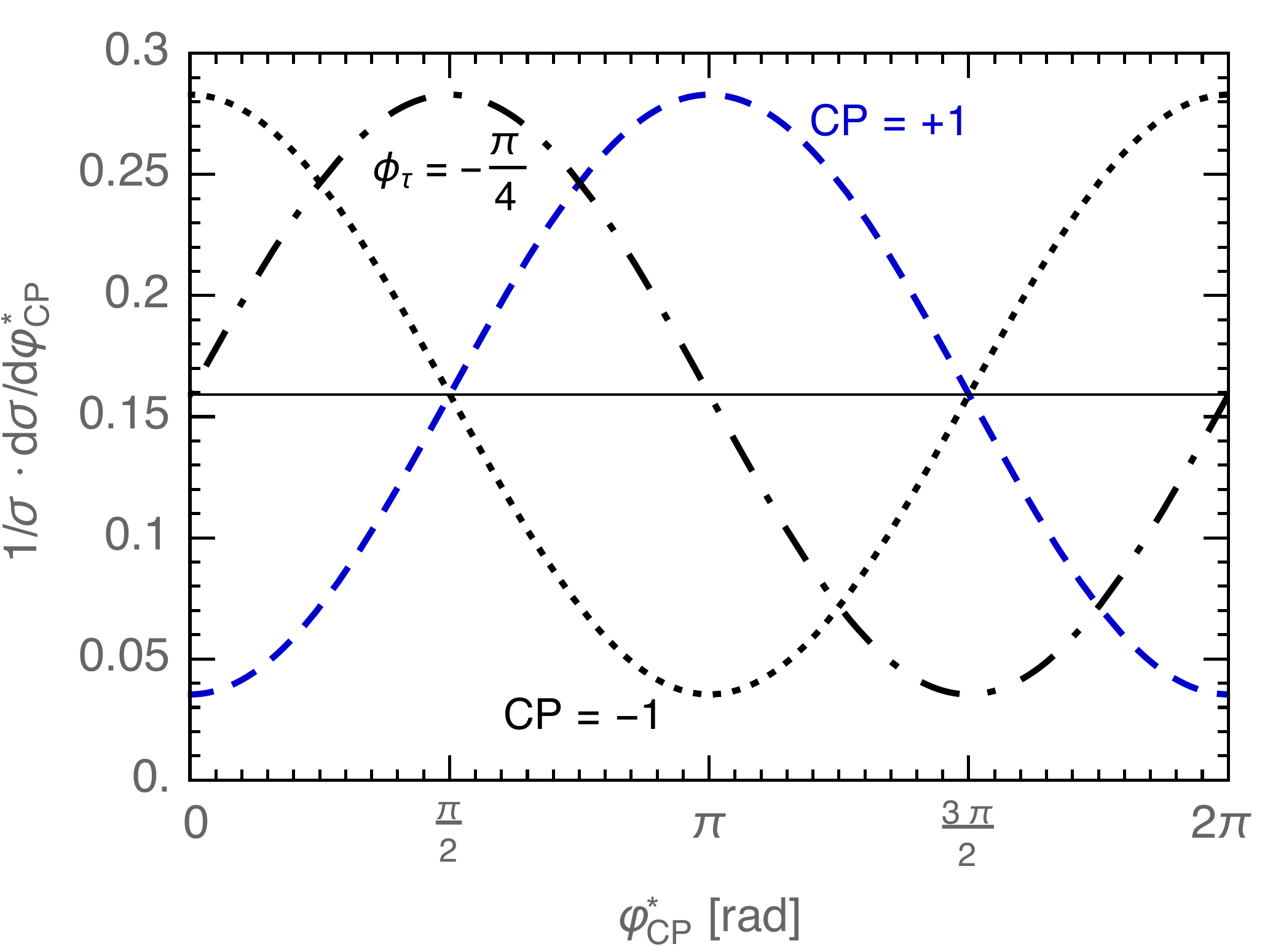}
\protect\protect\protect\caption{Left: Definition of 3-vectors and 
 the angle $\varphi_{CP}^{*}$ in the $a^{-}a'^{+}$
ZMF (here $\pi^{-}\pi^{+}$ ZMF) for the impact parameter method. Right: 
 Normalized $\varphi_{CP}^{*}$
distribution for  $pp\to h\to\tau^{-}\tau^{+}\to\pi^{-}\pi^{+}+2\nu_{\tau}$
at the LHC (13 TeV) at NLO QCD with $q_{T}^{\pi\pm}\ge20$~GeV and $|\eta_{\pi\pm}|<2.5$, 
 and $m_{h}=125$~GeV. The blue dashed, the black dotted and black long-dashed line  shows
the distribution for a CP-even Higgs boson ($\phi_{\tau}=0$), a CP-odd Higgs boson ($\phi_{\tau}=\pm{\pi}/{2}$)
and a CP mixture $(\phi_{\tau}=-{\pi}/{4})$, respectively. The solid black flat line is the distribution due to 
 the $Z^*/\gamma^*\to\tau\tau$ background.
 \label{fig:h_pipi_impact_method} }
\end{figure}

The method described in ~\cite{Berge:2008dr,Berge:2011ij,Berge:2013jra,Berge:2014sra} can be used for
 all $\tau^\mp$ decay modes \eqref{eq:taul} - \eqref{eq:taupi}
 if the charged prongs prongs $a^{-}, a'^{+}$ have a non-vanishing impact parameter.
  This method requires the measurement  of the 4-momenta of $a^{-}$ and $a'^{+}$ and their impact
 parameters vectors ${\bf{n}}_{\mp}$ in the laboratory frame. The vectors ${\bf{n}}_{\mp}$
  begin at the Higgs-boson production vertex (which should be known with some precision 
   also along the beam direction, which we take to be the $z$ direction)
    and end perpendicular on the $a^{-}$ and $a'^{+}$ tracks. The corresponding unit vectors
     are denoted by ${\bf \hat{n}}_{\mp}$.
  The 4-momenta $q_{-}^\mu$, $q_{+}^\mu$ of  $a^{-}, a'^{+}$ and
the impact parameter 4-vectors defined by $n_{\mp}^{\mu}=(0,{\bf \hat{n}}_{\mp})$
are boosted into the $a^{-} a'^{+}$ ZMF. The variables in the $a^{-} a'^{+}$ ZMF
are denoted by  an asterisk, for instance, $q_{\mp}^{*\mu}$, $n_{\mp}^{*\mu}$. 
 An observable that is sensitive to the  $CP$ nature of the Higgs boson is 
  obtained as follows: We decompose ${\bf n}_{\mp}^{*}$ into their
  normalized components ${\hat{\bf n}}_{\textbar\textbar}^{*\mp}$
and ${\hat{\bf n}}_{\perp}^{*\mp}$ which are parallel and perpendicular
to the respective 3-momentum ${\bf q}_{-}^{*}$ and ${\bf q}_{+}^{*}$.
An unsigned angle $\varphi^{*}$ ($0\leq\varphi^{*}\leq\pi$) and
a $CP$-odd and $T$-odd triple correlation ${\cal O}_{CP}^{*}$ ($-1\le{\cal O}_{CP}^{*}\le1$)
can be defined by 
\begin{equation}
\varphi^{*}=\arccos({\bf \hat{n}}_{\perp}^{*+}\cdot{\bf \hat{n}}_{\perp}^{*-}) \, ,\qquad
\qquad{\cal O}_{CP}^{*}={\bf \hat{q}}_{-}^{*}\cdot({\bf \hat{n}}_{\perp}^{*+}\times{\bf \hat{n}}_{\perp}^{*-})\,,\label{phistar}
\end{equation}
where ${\bf \hat{q}}_{-}^{*}$ is the normalized $a^{-}$ momentum
in the $a^{-}a'^{+}$ ZMF. Using these two quantities one can define
a signed angle $\varphi_{CP}^{*}$~\cite{Berge:2014sra} between
the $\tau^{-}\to a^{-}$ and $\tau\to a'^{+}$ decay planes by 
\begin{equation}
\varphi_{CP}^{*}=\left\{ \begin{array}{ccc}
\varphi^{*} & {\rm if} & {\cal O}_{CP}^{*}\geq0\,,\\
2\pi-\varphi^{*} & {\rm if} & {\cal O}_{CP}^{*}<0\, ,
\end{array}\right.\label{phistar_CP}
\end{equation}
 and $0\le\varphi_{CP}^{*}\le2\pi$.
A sketch of the definition of $\varphi_{CP}^{*}$ in the $a^{-}a'^{+}$
ZMF is given in Fig.~\ref{fig:h_pipi_impact_method}, left. 
 The distributions of $\varphi_{CP}^{*}$ were computed for inclusive Higgs production
 $ij\to hX$ (where $ij$ denote partons) and subsequent decays 
  $h\to\tau^-\tau^+\to a^{-}a'^{+}$ in \cite{Berge:2014sra}.
   The differential partonic cross section ${\hat\sigma}_{ij}$, integrated over the polar 
    angles of the charged prongs, is proportional to \\
     $1-\pi^2 b(E_-)b(E_+)\cos(\varphi_{CP}^{*}-2\phi_\tau)/16$, 
      where the functions $b(E_-)$, $b(E_+)$ defined in \eqref{eq:dGamma_dEdOmega} contain the information on the 
       $\tau$-spin analyzing power of $a^{-}$ and $a'^{+}$, respectively. From this distribution
        the Higgs mixing angle $\phi_\tau$ can be extracted. 
   
   For the computation of the  $\varphi_{CP}^{*}$ distributions we use the Monte Carlo program
    MCFM \cite{Campbell:2010ff}  to generate Higgs-boson production
  by  gluon-gluon fusion at NLO QCD. Using the narrow width approximation we include $h\to\tau^-\tau^+$ with
   $\tau$ spin correlations and the subsequent decays 
   $\tau^-\tau^+\to a^{-}a'^{+}$ with our own Monte Carlo code.
    As an example, we show in Fig.~\ref{fig:h_pipi_impact_method},
right, the normalized $\varphi_{CP}^{*}$ distribution
$pp\to h\to\tau^{-}\tau^{+}\to\pi^{-}\pi^{+}+2\nu_{\tau}$
for the LHC for a CP-even and CP-odd Higgs boson and for a CP-mixture.

 A possible non-vanishing scalar-pseudoscalar mixing angle $\phi_{\tau}$ can be extracted
from the shift of the measured distribution with respect to the SM
prediction (CP-even $h$, blue dashed line). One can determine $\phi_{\tau}$ 
 by fitting the function $f=u\cos\left(\varphi_{CP}^{*}-2\phi_\tau\right)+{\rm w}$
 to the measured        $\varphi_{CP}^{*}$  distributions
 for the respective final states $aa'$. The function is constrained by 
 $\int_{0}^{2\pi}d\varphi_{CP}^{*}f\,=\,2\pi{\rm w}\,=\,\sigma_{aa'}$,
where $\sigma_{aa'}$ is the $h$-production cross section including
the respective decay branching fractions. 
 The estimate of the  uncertainty of $\varphi_{CP}^{*}$
  for a given final state
depends on the values of the associated parameters $u$ and ${\rm w}$.
 For the comparison of different channels it is
 convenient to use the following asymmetry \cite{Berge:2013jra,Berge:2014sra}:
\begin{eqnarray}
A^{aa'}\; & =\frac{1}{\sigma_{aa'}}\,\int_{0}^{2\pi}\!\! d\varphi_{CP}^{*}
\left\{ d\sigma_{aa'}(u\cos(\varphi_{CP}^{*}-2\phi_\tau)>0)-d\sigma_{aa'}(u\cos(\varphi_{CP}^{*}-2\phi_\tau)<0)\right\} \nonumber \\
 & \,=\,\displaystyle{\frac{-4u}{2\pi{\rm w}}  }\,.
 \label{phiCP_asym}
\end{eqnarray}
 The values
of $A^{aa'}$ are independent of the mixing angle $\phi_\tau$ but do depend
on the product of the $\tau$-spin analyzing powers of $a$ and $a'$.
The larger $A^{aa'}$ the smaller the error $\Delta\phi_\tau$
in this decay channel, for a given number of events. The $\tau$-spin
analyzing power, and thus $A^{aa'}$, is maximal for the direct decays
$\tau^{\mp}\to\pi^{\mp}$ and for $\tau^{\mp}\to a_{1}^{L,T\mp}.$
In case of  the decays $\tau^{\mp}\to l^{\mp}$, $\tau^{\mp}\to\rho^{\mp}\to\pi^{\mp}$,
 and $\tau^{\mp}\to a_{1}^{\mp}\to\pi^{\mp}$ the $\tau$-spin analyzing power of the charged lepton, respectively
  of the charged pion can be enhanced by applying an appropriate cut on the energy of $l^{\mp}$ and 
  $\pi^{\mp}$, respectively.

For the  $\tau\tau\to \pi \pi$   decay channel, the asymmetry 
$A^{\pi\pi}=39\%$ if no cuts on the pions are applied.
While a cut on the rapidities of the charged pions does not change the
normalized $\varphi_{CP}^{*}$ distribution, rejecting pions with low $q_{T}$ 
increases the amplitude $u$. Applying the cuts $q_{T}^{\pi\pm}\ge 20$~GeV,
and $|\eta_{\pi\pm}|\le2.5$ on the final charged pions, 
 as was done in Fig.~\ref{fig:h_pipi_impact_method}, 
 increases the asymmetry to $A^{\pi\pi}=50\%$. 
 
 The asymmetry $A^{aa'}$ was computed in \cite{Berge:2014sra} 
  for all combinations of the $\tau$ decay modes~\eqref{eq:taul}~-~\eqref{eq:taupi} with appropriate cuts.
 An important feature of the $\varphi_{CP}^{*}$ distribution is 
  that the contribution from  the irreducible Drell-Yan background
     $Z^*/\gamma^*\to\tau\tau$  is flat for 
      all charged prongs $a,a'$, as shown in \cite{Berge:2014sra}.
 The Drell-Yan  contribution decreases
the height of the normalized distribution and thus the magnitude of
the asymmetry \eqref{phiCP_asym}, but is not a major obstacle 
 in extracting the Higgs mixing angle $\phi_\tau$.

\subsection{Method using the $\rho$-decay plane}
\label{suec:rhodecpl}
 For Higgs-boson production in $e^+e^-$ collisions and the subsequent
  decay channel \linebreak $h\to \tau^{-}\tau^{+}\to\rho^{-}\rho^{+}+ 2\nu_{\tau}$,
 a slightly different method was proposed and analyzed in Refs.~\cite{Bower:2002zx,Desch:2003mw,Desch:2003rw,Was:2002gv,Worek:2003zp}
 for determining the scalar-pseudoscalar mixing angle $\phi_{\tau}$. 
   This method requires that the tracks of
  the charged and neutral pion of each $\rho$ decay can be separated. That is, 
   both the charged and the neutral pion momenta  must be measured and correctly assigned
     to $\rho^\mp$.
    The charged and neutral pion momenta are
 then boosted into the $\rho^{-}\rho^{+}$ ZMF, and the resulting
  $\pi^-, \pi^0$ and $\pi^+, \pi^0$ 3-momenta in this frame define
  two decay planes. The angle between these  planes serves as discriminating variable
  for determining the CP nature of $h$. This approach was applied in the recent studies
   \cite{Harnik:2013aja,Przedzinski:2014pla,Askew:2015mda} for Higgs-boson production at the LHC.

\begin{figure}[tb]
\noindent \centering{}\includegraphics[height=5.7cm]{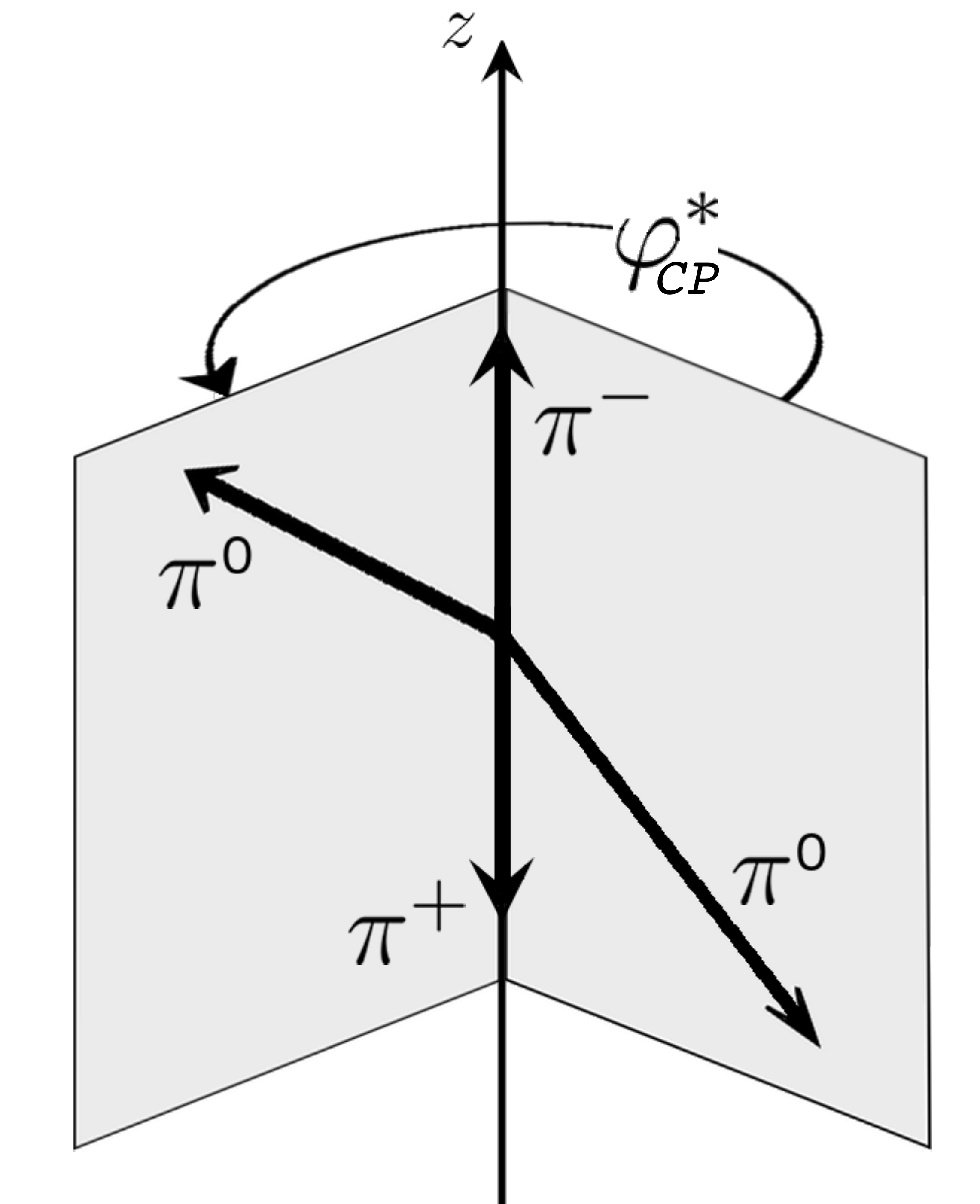}
\protect\protect\caption{Illustration of $\varphi_{CP}^{*}$ in 
 the $\rho$ decay-plane method as defined in \eqref{eq:Def_varphi_star_rho_rho}
 for   $pp\to h^{0}\to\tau^{-}\tau^{+}\to\rho^{-}\rho^{+}+2\nu$.
 \label{fig:rhorho_definition_of_phistarCP} }
\end{figure}

\begin{figure}[t]
\includegraphics[height=5.15cm]{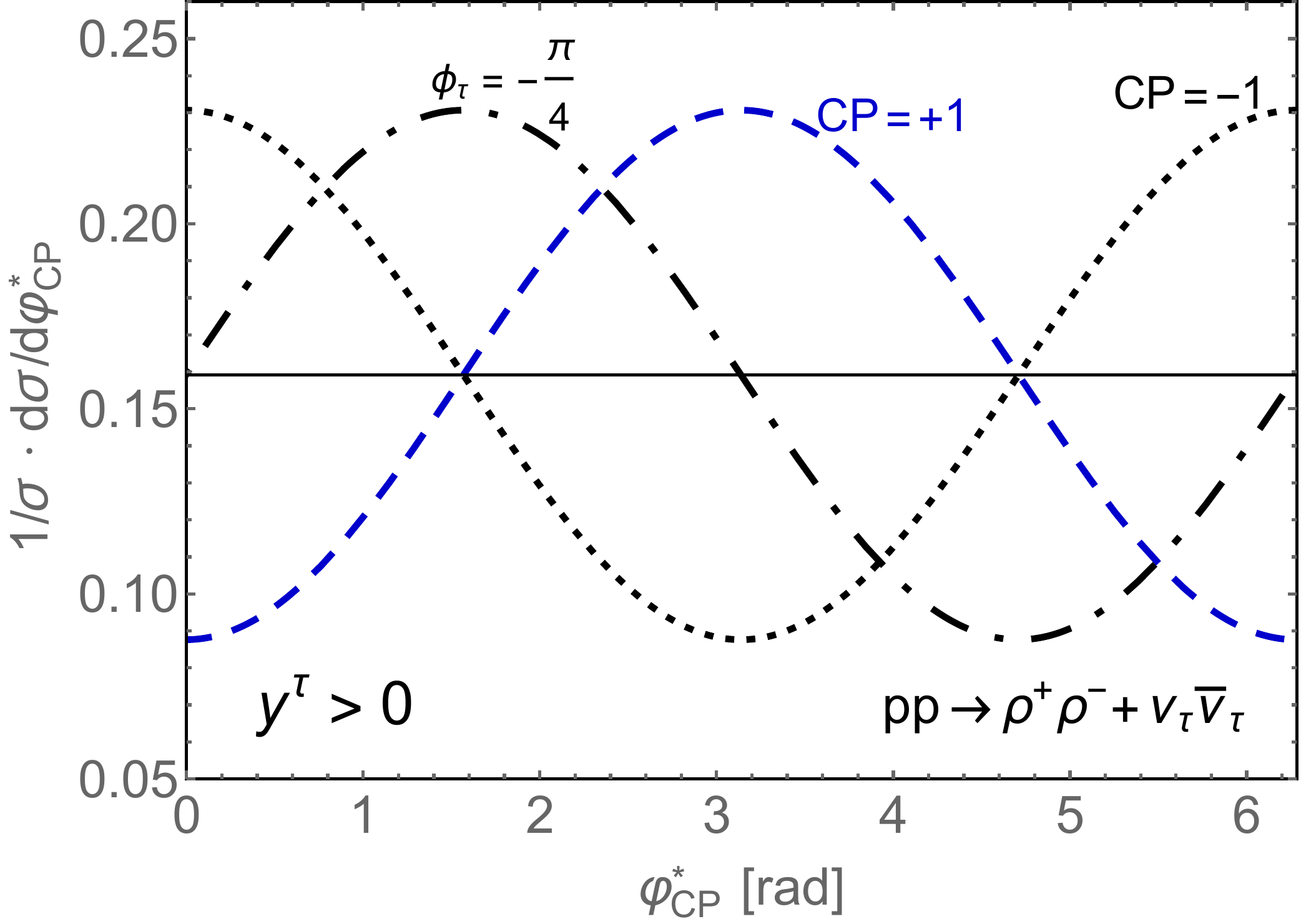}\hspace{0.5cm}
\includegraphics[height=5.15cm]{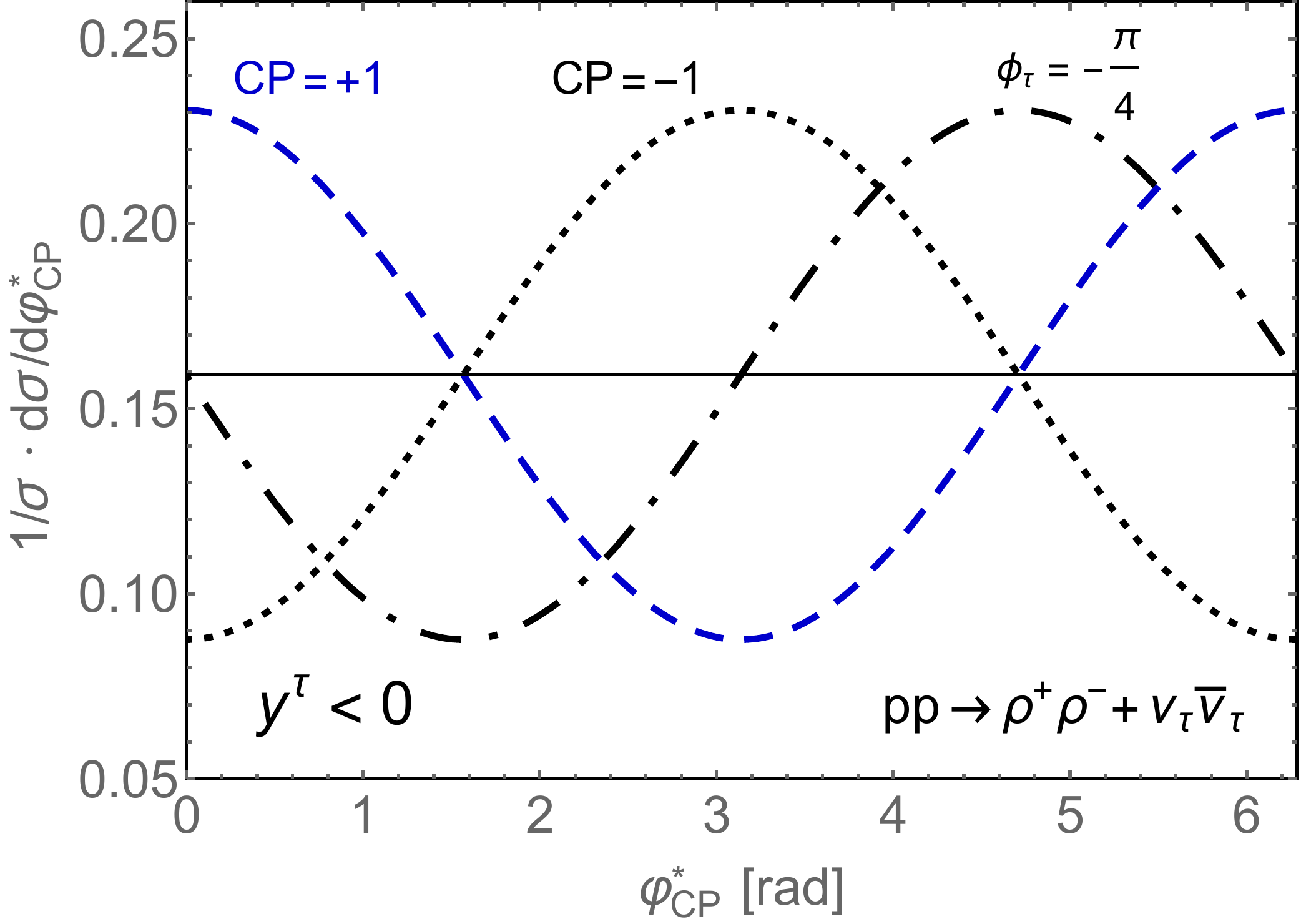}
\protect\protect\protect\caption{Normalized $\varphi_{CP}^{*}$ distribution
 obtained with the $\rho$-decay plane method \eqref{eq:Def_varphi_star_rho_rho}
for $pp\to h \to \tau^{-}\tau^{+}\to\rho^{-}\rho^{+}+2\nu_{\tau}$ at the
LHC (13 TeV).  The left and right plots are for events with $y^{\tau}>0$ and $y^{\tau}<0$, as defined in 
 \eqref{eq:Def_y_tau-restframe}.
 The cuts $p_{T}^{\rho\pm}\ge20$~GeV and $|\eta_{\rho\pm}|, |\eta_{\pi\pm}|\le2.5$ were
 applied.  
The blue dashed, black dotted, and black
long-dashed lines  show the distribution for a $CP$-even Higgs boson ($\phi_{\tau}=0$), a $CP$-odd Higgs boson $(\phi_{\tau}=\pm{\pi}/{2}$), and 
 a $CP$ mixture $(\phi_{\tau}=-{\pi}/{4})$, respectively. The flat lines are the distributions due to  Drell-Yan production.
 \label{fig:h_rhorho_ptrho25} }
\end{figure}

     Rather than choosing the $\rho^{-}\rho^{+}$ ZMF we use in the following, as for the impact parameter method,
   the  $a^{-}a'^{+}$ ZMF of the charged pions from $\rho^\mp$ decay. This allows us to standardize the definition
    of the discriminating variable for both methods. In the remainder of this section we define this variable
     for the $h\to \tau^{-}\tau^{+}\to\rho^{-}\rho^{+}$ decay channel. One boosts the 
      $\pi^-, \pi^0$ and $\pi^+, \pi^0$ 4-momenta, measured in the laboratory frame, into the $\pi^-\pi^+$ ZMF.
      In this frame, we denote the $\pi^-, \pi^0$  $(\pi^+, \pi^0)$ 3-momenta by 
     ${\bf q}^{*-}, {\bf q}^{*0-}$ $({\bf q}^{*+}, {\bf q}^{*0+})$. In the  $\pi^-\pi^+$ ZMF we
      compute, for each neutral pion, the normalized vector
       $\hat{\bf q}_{\perp}^{*0-}$ and $\hat{\bf q}_{\perp}^{*0+}$ which is transverse 
      to the direction of the associated charged pion.  The
  angle between these two vectors is given by\footnote{In \eqref{eq:defphidecp}
  and in \eqref{eq:Def_varphi_star_rho_rho} we use the same notation as in \eqref{phistar}
  and \eqref{phistar_CP}, respectively, in order not to overload the notation. }
\begin{align}
\varphi^{\,*}\,\,=\,\,\arccos\left(\hat{\bf q}_{\perp}^{*0+}\cdot \hat{\bf q}_{\perp}^{*0-}\right) \, , & \qquad0\le\varphi^{\,*}\le\pi \, .
\label{eq:defphidecp}
\end{align}
 In order to define a signed angle we use the 
   $CP$-odd triple correlation ${\cal O}^{*}$
\begin{align*}
{\cal O}^{*}\,\,=\,\,\hat{\bf q}^{*-}\cdot\left(\hat{\bf q}_{\perp}^{*0+}\times\hat{\bf q}_{\perp}^{*0-}\right) \, , & \qquad-1\le{\cal O}^{*}\le+1 \, . 
\end{align*}
The discriminating variable that is sensitive to the mixing angle $\phi_{\tau}$ is defined by
\begin{equation}
\varphi_{CP}^{*}=\left\{ \begin{array}{ccc}
\varphi^{\,*} & {\rm if} & {\cal O}^{*}\geq0\,\\
2\pi-\varphi^{\,*} & {\rm if} & {\cal O}^{*}<0\,
\end{array}\right.,\quad\quad{\rm with}\,\,\quad0\le\varphi_{CP}^{*}\le2\pi\,.
\label{eq:Def_varphi_star_rho_rho}
\end{equation}
The angle $\varphi_{CP}^{*}$ is shown in Fig.~\ref{fig:rhorho_definition_of_phistarCP}.
 In order to  obtain a non-trivial $\varphi_{CP}^{*}$ distribution,
  one needs to separate the events into two classes depending
 on the sign of the $\tau^{\mp}$ spin-analyzing functions or polarimeter vectors
  associated with the  $\tau^{\mp}\to \rho^{\mp}$ decays  \cite{Bower:2002zx,Desch:2003mw,Desch:2003rw,Was:2002gv,Worek:2003zp}.
  These polarimeter vectors are proportional
to $E_{\pi^{\pm}}-E_{\pi^{0}}$,
where $E_{\pi^{\pm}}$ and $E_{\pi^{0}}$ are the energies of the pions associated with
 the decays of $\rho^\pm$  in the  respective
$\tau^{\pm}$ rest frames. Using the variables 
 \begin{eqnarray}
y_{-}^{\tau}\,\,=\,\,\frac{\left(E_{\pi^{-}}-E_{\pi^{0}}\right)}{\left(E_{\pi^{-}}+E_{\pi^{0}}\right)} & \qquad{\rm and}\qquad & y_{+}^{\tau}
 \,\,=\,\,\frac{\left(E_{\pi^{+}}-E_{\pi^{0}}\right)}{\left(E_{\pi^{+}}+E_{\pi^{0}}\right)}
\label{eq:Def_y_tau-restframe}
 \end{eqnarray}
and selecting 
\begin{eqnarray}
y^{\tau}\,>\,0\qquad{\rm or}\qquad y^{\tau}\,<\,0 & \quad,\qquad\qquad & {\rm where}\,\,\,\, y^{\tau}\,\,=\,\, y^\tau_{-}y^\tau_{+} \, , 
 \label{eq:Def_y1_y2}
\end{eqnarray}
the events are divided into two classes.
  The resulting distributions are shown
in Fig.~\ref{fig:h_rhorho_ptrho25} for the gluon-gluon fusion process
$pp\to h \to\tau^{-}\tau^{+}\to\rho^{-}\rho^{+}+2\nu$  for the LHC (13 TeV).

The definition of $\varphi_{CP}^{*}$ in Eq.~\eqref{eq:Def_varphi_star_rho_rho}
has been chosen such that the resulting Higgs-boson distributions 
 in Fig.~\ref{fig:h_rhorho_ptrho25} left, i.e. for $y^{\tau}>0$,
   agree with the respective distribution obtained with the 
impact parameter method and with those\footnote{The distributions given 
in~\cite{Bower:2002zx,Desch:2003mw,Was:2002gv,Worek:2003zp} are
shifted by an angle of $\pi$ due to a different definition of $\varphi^{*}$
which uses normalized vectors perpendicular to the planes spanned
by the $\rho$ mesons and their decay products.}
 of Ref.~\cite{Desch:2003rw}.
 For $y^{\tau}\,<\,0$ (Fig.~\ref{fig:h_rhorho_ptrho25}, right)
 all $\varphi_{CP}^{*}$ distributions  are 
  shifted by $\varphi_{CP}^{*}\to\varphi_{CP}^{*}+\pi$, as compared
   to those of Fig.~\ref{fig:h_rhorho_ptrho25}, left.
   This is because in this case the product of the two $\tau^{\mp}\to \rho^{\mp}$
    spin-analyzing functions is negative. 
 
 The cuts on $y^{\tau}$ are academic because the $\tau$ rest frames can in general not be 
  reconstructed. Therefore, we use in the following the variables
\begin{eqnarray}
y_{-}^{\, L}\,\,=\,\,\frac{\left(E_{\pi^{-}}^{L}-E_{\pi^{0}}^{L}\right)}{\left(E_{\pi^{-}}^{L}+E_{\pi^{0}}^{L}\right)} 
& \qquad{\rm and}\qquad & y_{+}^{\, L}\,\,=\,\,\frac{\left(E_{\pi^{+}}^{L}-E_{\pi^{0}}^{L}\right)}{\left(E_{\pi^{+}}^{L}
 + E_{\pi^{0}}^{L}\right)} \, , 
\label{eq:Def_y1_y2_labframe}
\end{eqnarray}
where $E_{\pi^{\pm}}^{\, L}$ and $E_{\pi^{0}}^{\, L}$ 
energies of the charged and the neutral pions associated with
 the decays of $\rho^\pm$ in the laboratory frame.
Again, using $y^{L}=y_{-}^{L}y_{+}^{L}$
 one separates events into two classes according to $y^{L} > 0$
  and $y^{L}<0.$

\begin{figure}[tb]
\noindent \centering{}
 \includegraphics[height=5.7cm]{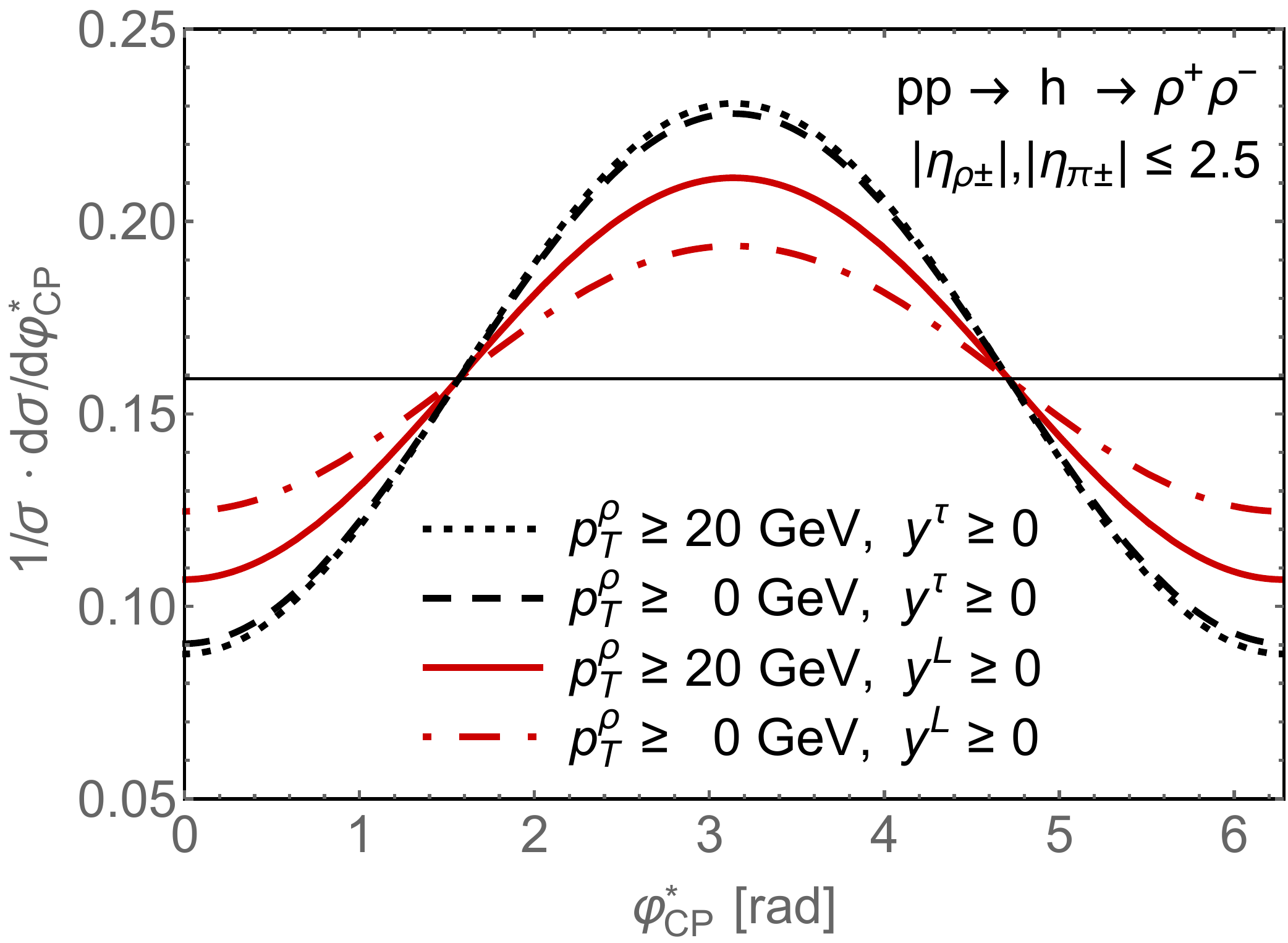}
\protect\protect\caption{$\varphi_{CP}^{*}$ distribution  defined in 
 Eq.~\eqref{eq:Def_varphi_star_rho_rho} for 
  $p p\to h \to \tau^{-}\tau^{+}\to \rho^{-}\rho^{+}+2\nu$
at LHC (13 TeV) for a CP-even Higgs boson using the $\rho$-decay plane method.
The black dotted and  black dashed line  (solid red line and  dot-dashed red line) 
 show the distribution if the cut on $y=y_{1} y_{2}$ is performed in the corresponding
$\tau$ rest frames (in the laboratory frame). For the distributions displayed
  by the solid red line and black dotted line an additional cut  of $p_{T\rho}\ge 20$~GeV
   on the hadronic $\tau$-jet was applied.
\label{fig:h_rhorho_y_cuts} }
\end{figure}

 The effect of selecting events with respect to the sign of $y^{\tau}$ 
 and $y^{L}$ on the $\varphi_{CP}^{*}$ distribution are compared
  in Fig.~\ref{fig:h_rhorho_y_cuts} for the case of a CP-even Higgs boson.
   If no cuts on the transverse momenta of the $\rho$ mesons are applied
   the asymmetry  \eqref{phiCP_asym} associated with the curve $y^L\ge 0$ (red dot-dashed line, $A^{\rho\rho}\simeq 14\%$)
   is reduced considerably with respect to the one for $y^\tau\ge 0$ (black dashed line, $A^{\rho\rho}\simeq 28\%$).

However, the situation is different if a minimum $p_{T}$-cut on the 
 hadronic jet from $\tau\to \rho$ decay is imposed. We use $p_{T\rho}\ge20$~GeV.
At the LHC such a cut is indispensable in order to suppress the QCD background.
 In this case  the asymmetry \eqref{phiCP_asym} is much less reduced if
  one selects the events with respect to $y^{L}$ rather than $y^\tau$, namely from
  $A^{\rho\rho}\simeq 29\%$ to $A^{\rho\rho}\simeq 21\%$. 

%
\subsection{Combination of impact parameter and $\rho$-decay plane method}
\label{susec:combi}

In this section we combine the impact parameter and $\rho$-decay plane method
 of Sec.~\ref{susec:impactp} and~\ref{suec:rhodecpl} and 
  define the discriminating variable $\varphi_{CP}^{*}$ for this case.

Let us consider the decay channels $h\to\tau^-\tau^+\to a^- \rho^+$.
 For the $\tau^+\to \rho^+\to\pi^+\pi^0$ decay we assume that the momenta of the charged
  and neutral pion can be measured. 
 The $\tau^{-}$  may decay via one of the  major decay channels $\tau^{-}\to a^{-}$
 listed in Eqs.~\eqref{eq:taul} - \eqref{eq:taupi}, including the decay via a $\rho^{-}$
 meson\footnote{Here we assume that the momenta of the charged and neutral pion from $\rho^-$
  decay  can not be separated with sufficient precision. Otherwise
one would use the $\rho$-decay plane method as described in Sec.~\ref{suec:rhodecpl}.}.
 For the $\tau^{-}$ decay we demand a non-vanishing impact
parameter of the final charged prong $a^{-}$.

As before, the variable $\varphi_{CP}^{*}$ will be defined
 in the zero-momentum-frame of the final charged prongs $a^{-}a'^+$,
 where in the case at hand $a^{-}=e^-, \mu^-, \pi^-$ and $a'^+=\pi^+$.
 The $\pi^{+}$ and $\pi^{0}$ momenta from $\tau^{+}\to \rho^+$ decay
  are boosted into the $a^{-}a'^{\text{+ }}$ ZMF and we calculate
in this frame, as described in Sec.~\ref{suec:rhodecpl}, the
transverse  neutral-pion direction with respect to the charged
pion momentum. The resulting normalized 
 vectors are denoted by  $\hat{\bf q}_{\perp}^{*0+}$ and
  $\hat{\bf q}^{*+}$, respectively.
 For the  $\tau^{-}\to a^-$ decay we boost
 the 4-momentum of the charged prong and its corresponding impact parameter
  vector $n^{\mu-}=(0,{\hat{\bf n}}^{\,-})$ also into the $a^{-}a'^+$
ZMF.  The resulting 4-vectors are denoted
by $q^{\mu *-}$ and $n^{\mu *-}$. In the $a^{-}a'^+$ ZMF we
calculate the normalized transverse vector $\hat{\bf n}_{\perp}^{*-}$
  as described in in Sec.~\ref{susec:impactp}. 
  The normalized 3-momentum of the charged prong $a^{-}$ is denoted by $\hat{\bf q}^{*-}$.
   With these variables, an angle  $\varphi^{\,*}$ and 
    a triple correlation ${\cal O}^{*}$ are defined by
\begin{align}
\varphi^{\,*}\,\,= & \,\,\,\arccos\left(\hat{\bf q}_{\perp_{\,_{\,_{\,}}}}^{*0+}\cdot \hat{\bf n}_{\perp}^{*-}\right) \, , \qquad
\qquad{\cal O}^{*}\,\,=\,\,\,\hat{\bf q}^{*-}\cdot\left(\hat{\bf q}_{\perp}^{*0+}\times\hat{\bf n}_{\perp}^{*-}\right)\, ,
\end{align}
 and the resulting variable which is sensitive to the CP nature of $h$ is, as before, 
\begin{equation}
\varphi_{CP}^{*}=\left\{ \begin{array}{ccc}
\varphi^{\,*} & {\rm if} & {\cal O}^{*}\geq0\,\\
2\pi-\varphi^{\,*} & {\rm if} & {\cal O}^{*}<0\,
\end{array}\right.,\quad\quad{\rm with}\,\,\quad 
 0\le\varphi_{CP}^{*}\le2\pi\, .
 \label{eq:Def_varphi_star_combined}
\end{equation}
In order to obtain a non-trivial $\varphi_{CP}^{*}$
 distribution one has to separate, as described in Sec.~\ref{suec:rhodecpl}, 
   events from $\tau^{+}\to\rho^{+}$ decay which have positive and negative values of $y^L_+$ 
    defined in \eqref{eq:Def_y1_y2_labframe}.
 Also the $\tau^{-}\to a^{-}$ events may have to be divided into two classes, depending
  on the decay mode, cf. \cite{Berge:2014sra}. 
For the direct decay $\tau^{-}\to\pi^{-}+\nu_{\tau}$, for $\tau^{-}\to a_1^{-L,T} +\nu_{\tau}$, and for the leptonic
decays $\tau^{-}\to l^{-}+\bar{\nu}_{l}+\nu_{\tau}$ such a separation
is, however, not necessary.

 For the decays $h\to\tau^-\tau^+\to \rho^- a^+$ one proceeds analogously to
  the charge-conjugate modes described above. In this case 
 the angle $\varphi^{\,*}$ and ${\cal O}^{*}$ are defined by
\begin{align}
\varphi^{\,*}\,\,= & \,\,\,\arccos\left(\hat{\bf q}_{\perp}^{*0-}\cdot\hat{\bf n}_{\perp}^{*+}\right) \, , \qquad
\qquad{\cal O}^{*}\,\,=\,\,\,\hat{\bf q}^{*-}\cdot\left( \hat{\bf n}_{\perp}^{*+}\times \hat{\bf q}_{\perp}^{*0-} \right)\, , 
\end{align}
 and $\varphi_{CP}^{*}$ is given again by \eqref{eq:Def_varphi_star_combined}.
 Here , the events from $\tau^{-}\to\rho^{-}$ decay which have positive and negative values of $y^L_-$ 
 (cf. \eqref{eq:Def_y1_y2_labframe}) have to be separated. In addition, also
 the  $\tau^{+}\to a^{+}$ may have to be divided into two classes, see  \cite{Berge:2014sra}.

\begin{figure}[t]
\noindent \raggedright{}\hspace*{1cm}
\includegraphics[height=5.7cm]{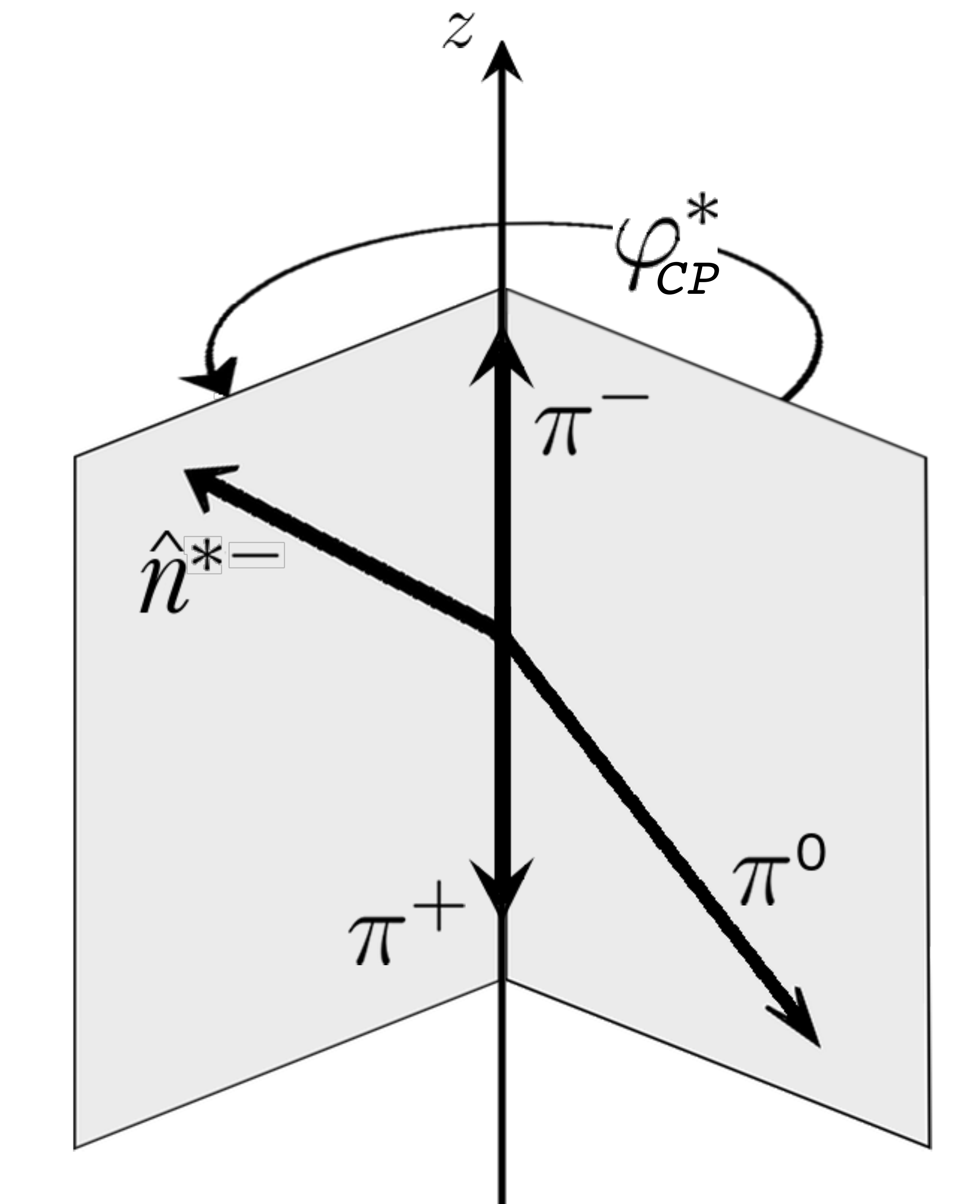}\hspace*{0cm}\vspace{-5.3cm} \\
 \hspace*{6.7cm}
 \includegraphics[height=5.7cm]{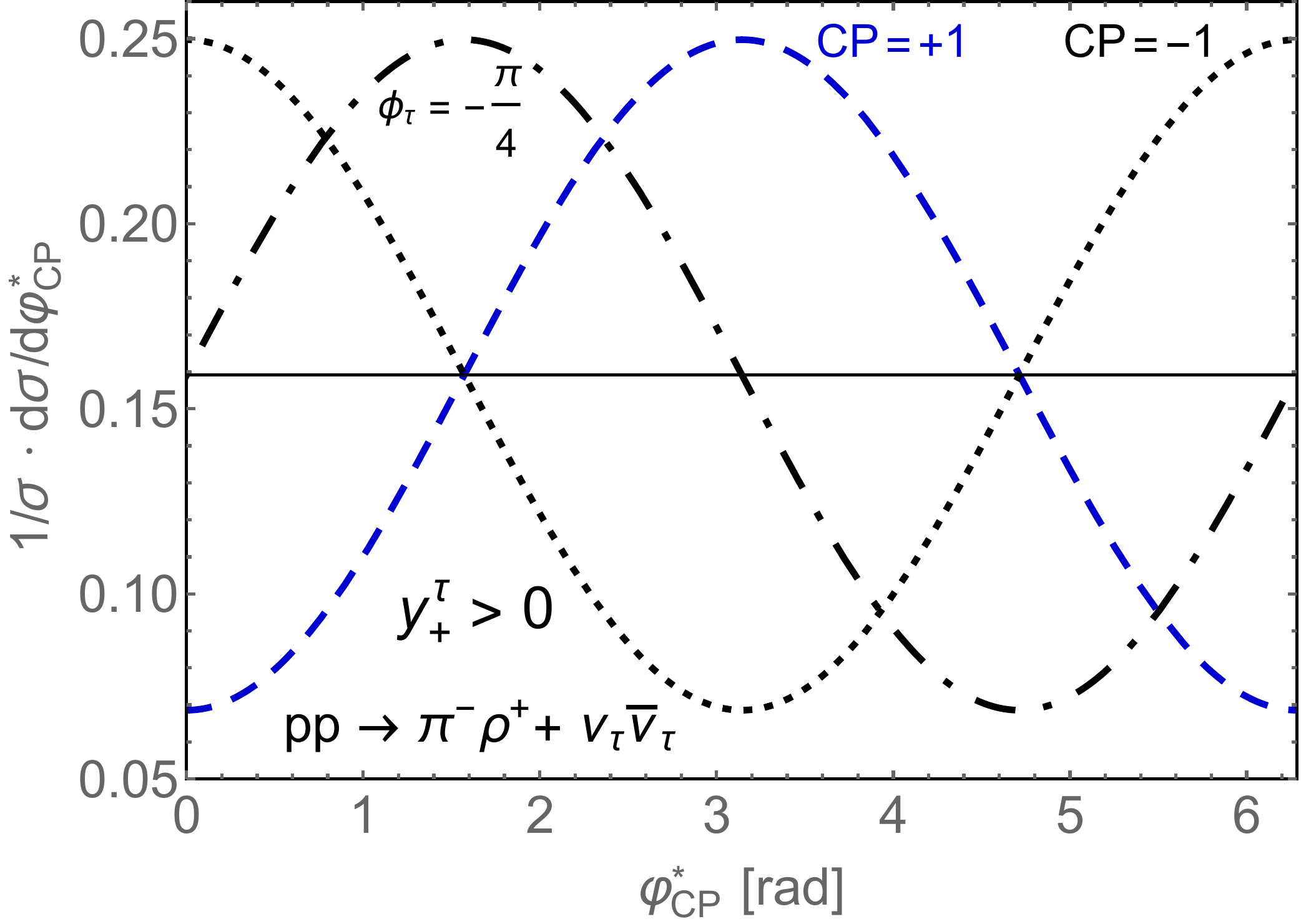}
\protect\protect\protect\caption{Left: Definition of the angle $\varphi_{CP}^{*}$ in case of the combined
method. Right: Normalized $\varphi_{CP}^{*}$ distributions, Eq.~\eqref{eq:Def_varphi_star_combined},
 for $pp\to h \to \tau^{-}\tau^{+}\to\pi^{-}\rho^{+}+2\nu_{\tau}$ events  with $y_{+}^{\tau}>0$
at the LHC (13 TeV). The cuts $q_{T}^{\pi^-},p_{T}^{\rho^+}\ge20$~GeV  and
 $|\eta_{\rho^{+}}|, |\eta_{\pi^\pm}|\le2.5$ were applied.
The blue dashed, black dotted, and black
long-dashed lines  show the distribution for a $CP$-even Higgs boson ($\phi_{\tau}=0$), a $CP$-odd Higgs boson $(\phi_{\tau}=\pm{\pi}/{2}$), and 
 a $CP$ mixture $(\phi_{\tau}=-{\pi}/{4})$, respectively. The flat line is the distribution due to  Drell-Yan production.
  The distributions for events with $y_{+}^{\tau}<0$ are shifted by $\varphi_{CP}^{*} \to \varphi_{CP}^{*} + \pi$.
\label{fig:hacpmix_impactrho_ytau+gt0_ytau+lt0} }
\end{figure}

For the decays $h\to\tau^-\tau^+\to a^- \rho^+$ the definition of the 
 angle $\varphi_{CP}^{*}$ is sketched in Fig.~\ref{fig:hacpmix_impactrho_ytau+gt0_ytau+lt0},
left. For the specific case where $\tau^-$ decays directly to $\pi^-$, 
 the $\varphi_{CP}^{*}$ distribution is shown for events with $y_{+}^{\tau}>0$
 in Fig.~\ref{fig:hacpmix_impactrho_ytau+gt0_ytau+lt0},
right. For demonstration purposes we used here  the variable $y_{+}^{\tau}>0$
 defined in \eqref{eq:Def_y_tau-restframe} rather than  $y^L_+$. Cuts as given in the caption
  of this figure are applied. The asymmetry \eqref{phiCP_asym} computed from these distributions is $A=36\,\%$. 
   This asymmetry
is somewhat larger than the asymmetry associated with the decay
$h\to \tau^-\tau^+\to\rho^{-}\rho^{+}$ (cf. Fig.~\ref{fig:h_rhorho_ptrho25}), because the 
 $\tau$-spin analyzing
power of the direct decay $\tau^{-}\to\pi^-$ is one.
The distributions for $y_{+}^{\tau}<0$, which are not shown,
  are shifted by  $\varphi_{CP}^{*} \to \varphi_{CP}^{*} + \pi$.

The $\varphi_{CP}^{*}$ distributions for the reactions with charge-conjugate final states,
$pp\to h \to \tau^{-}\tau^{+}\to\rho^{-}\pi^{+}+2\nu_{\tau}$,
 are the same  as those shown in Fig.~\ref{fig:hacpmix_impactrho_ytau+gt0_ytau+lt0}, right,
  for events with $y_{-}^{\tau}>0$ and if cuts analogous to those given in the caption of this
   figure are applied.
  The distributions for events with $y_{-}^{\tau}<0$ are again shifted by
 $\varphi_{CP}^{*} \to \varphi_{CP}^{*} + \pi$.

\begin{figure}[t]
\includegraphics[height=5.15cm]{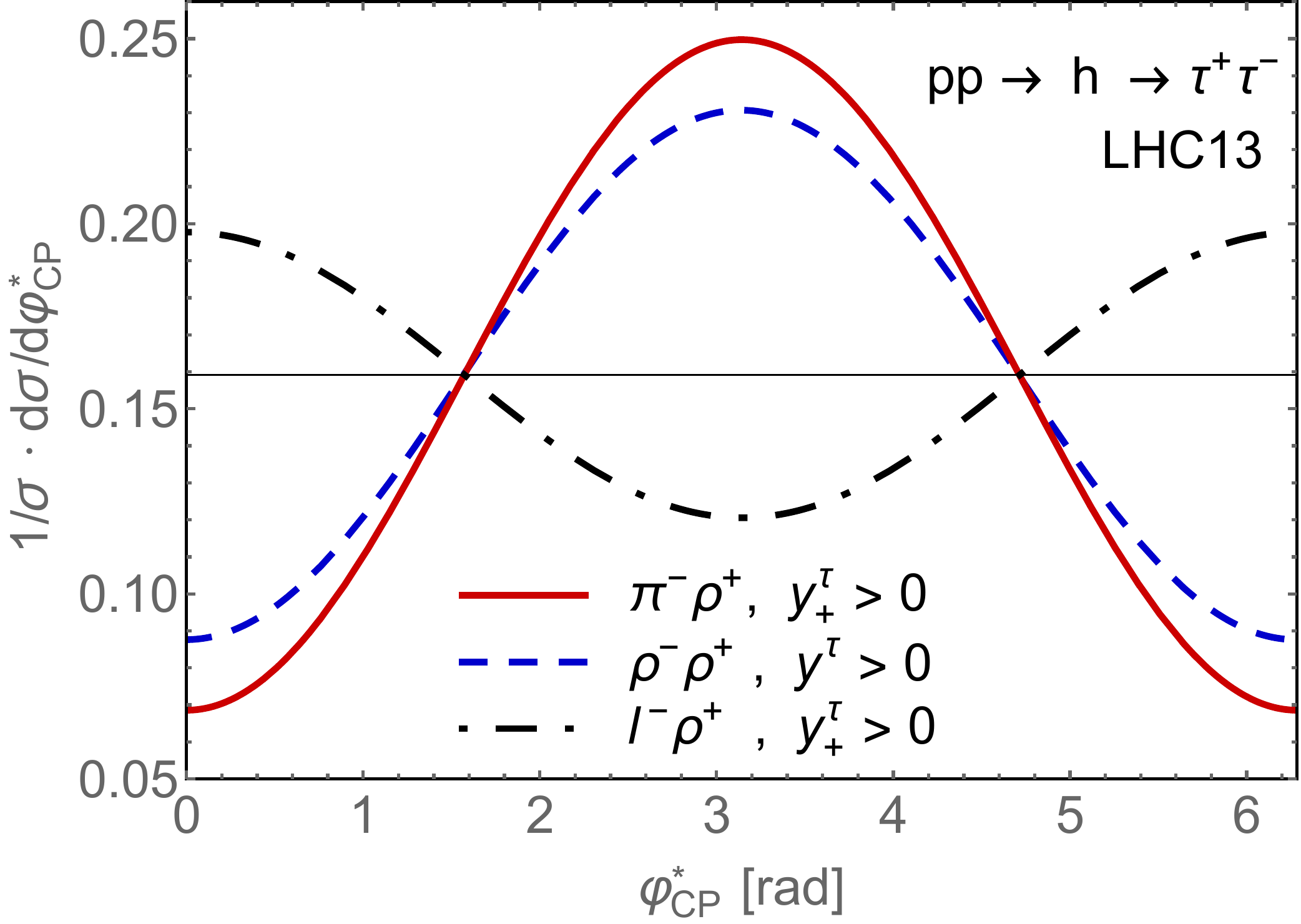}\hspace{0.5cm}
\includegraphics[height=5.15cm]{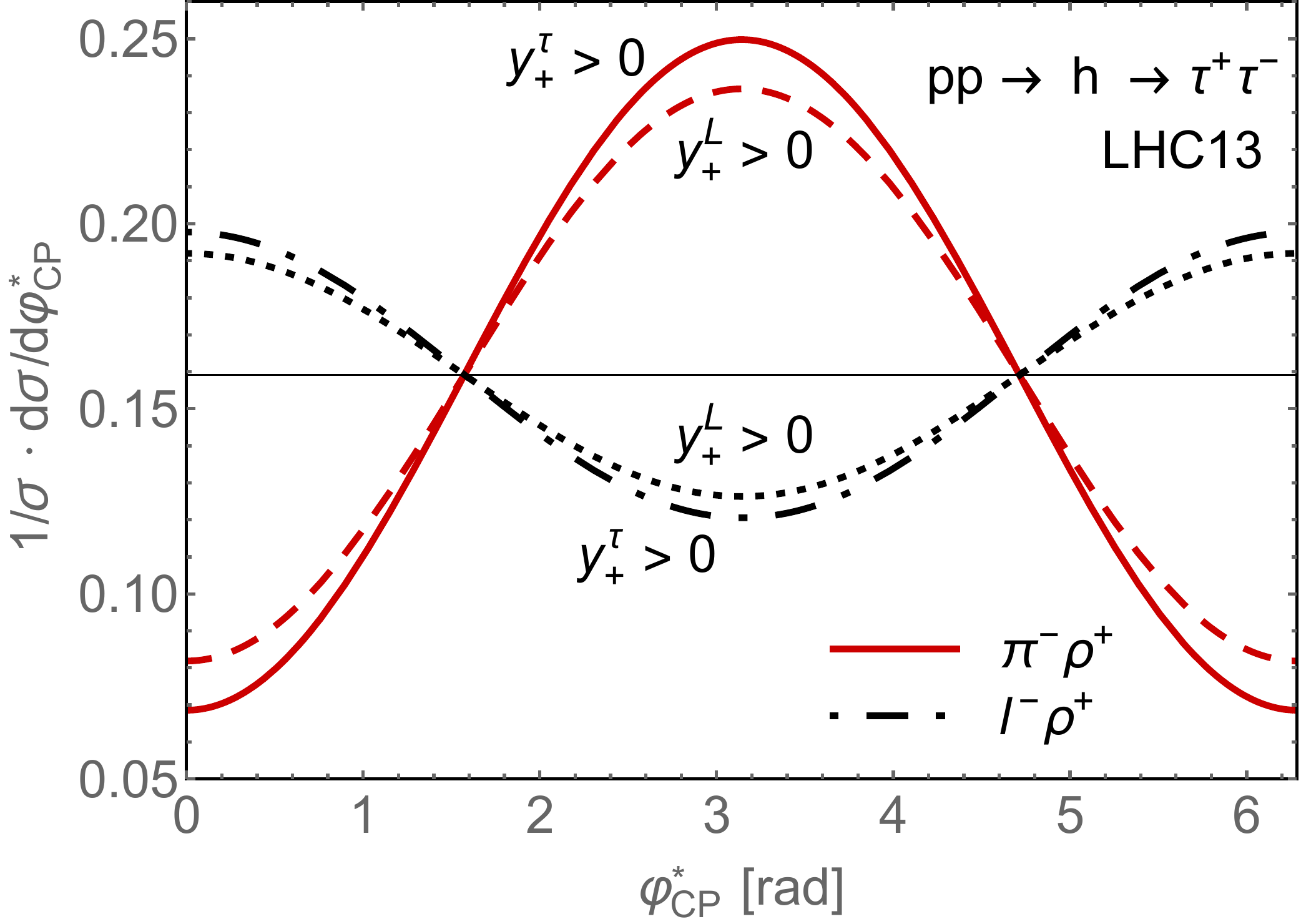}
\protect\protect\protect\caption{Normalized 
 $\varphi_{CP}^{*}$ distributions for a  CP-even Higgs boson 
at the LHC (13 TeV). The cuts  $p_{T}^{\rho^\pm}\ge20$~GeV,
 $|\eta_{\rho^\pm}|, |\eta_{\pi^\pm}|, |\eta_{l^-}|\le 2.5$ were applied.
 The dashed blue line (left plot) is the distribution for $h \to\rho^{-}\rho^{+}$,
as in Fig.~\ref{fig:h_rhorho_ptrho25}, left. The solid red and dashed
red lines (left and right plot) correspond to   $h\to\pi^{-}\rho^{+}$ with the additional cut $q_{T}^{\pi^-}\ge 20$~GeV.
 The dot-dashed black and dotted black lines (left and right plots) 
 correspond to  $h\to l^{-}\rho^{+}$ with the additional cut $q_{T}^{l^-}\ge 20$~GeV.
  The distributions shown in the left plot 
 were computed with the  cuts $y^{\tau}>0$ and $y_{+}^{\tau}>0$.
  The plot on the right shows
       the reduction of the height of the distributions 
 if the laboratory-frame cut $y_{+}^{L}>0$ is applied. \label{fig:h_impactrho_y_cuts} }
\end{figure}

Let us consider, for definiteness, the decay of a CP-even Higgs boson.
Fig.~\ref{fig:h_impactrho_y_cuts}, left, shows 
the $\varphi_{CP}^{*}$ distribution for $h\to \pi^{-}\rho^{+}$ obtained with
 the combined method (solid red line), while the  dashed blue line
  is the distribution for $h\to \rho^{-}\rho^{+}$ obtained with
   the $\rho$-decay plane method.
 For $h\to \pi^{-}\rho^{+}$ the height of the distribution 
is somewhat larger than for $h\to \rho^{-}\rho^{+}$.
 This demonstrates the importance of using the combined method
  for eventually attaining a good precision of the
   Higgs mixing angle $\phi_{\tau}$. The $\varphi_{CP}^{*}$ distribution (dot-dashed black line)
    for $h\to l^{-}\rho^{+}$ is also shown in this figure,
   This distribution has its minimum
at $\varphi_{CP}^{*}=\pi$ because the $\tau$-spin analyzing function of 
$l^{-}$ is negative.

The plots in  Fig.~\ref{fig:h_impactrho_y_cuts}, right, demonstrate that
 a realistic event selection, i.e., applying the laboratory-frame cut $y_{+}^{L}>0$,
  does not significantly change the shape of the distribution.
 Because here only one $\rho$ meson is involved, the effect of such a cut is much smaller than
for~$h \to \rho^{+}\rho^{-}$.

\begin{figure}[t]
\includegraphics[height=5.15cm]{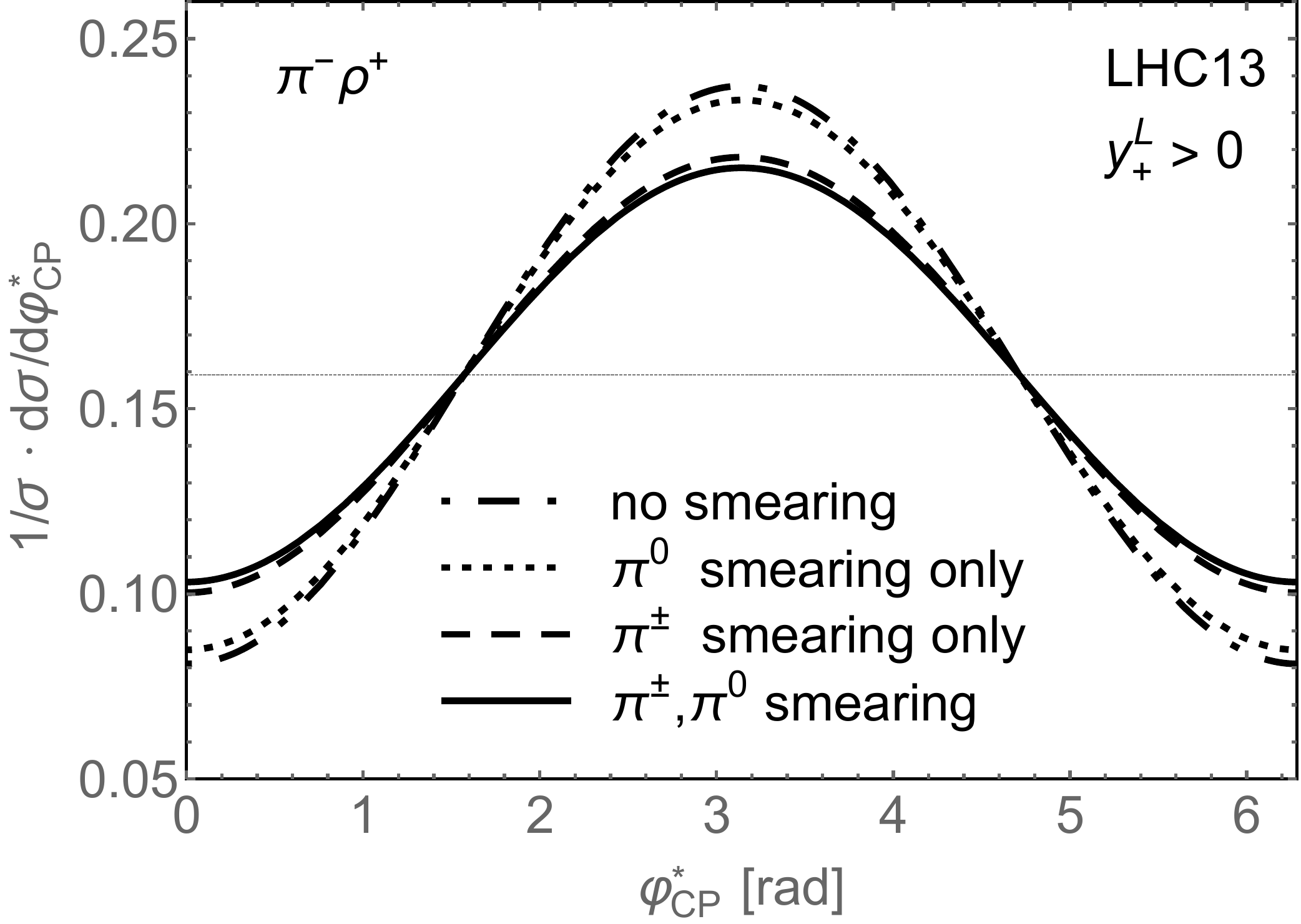}\hspace*{3mm}
\includegraphics[height=5.15cm]{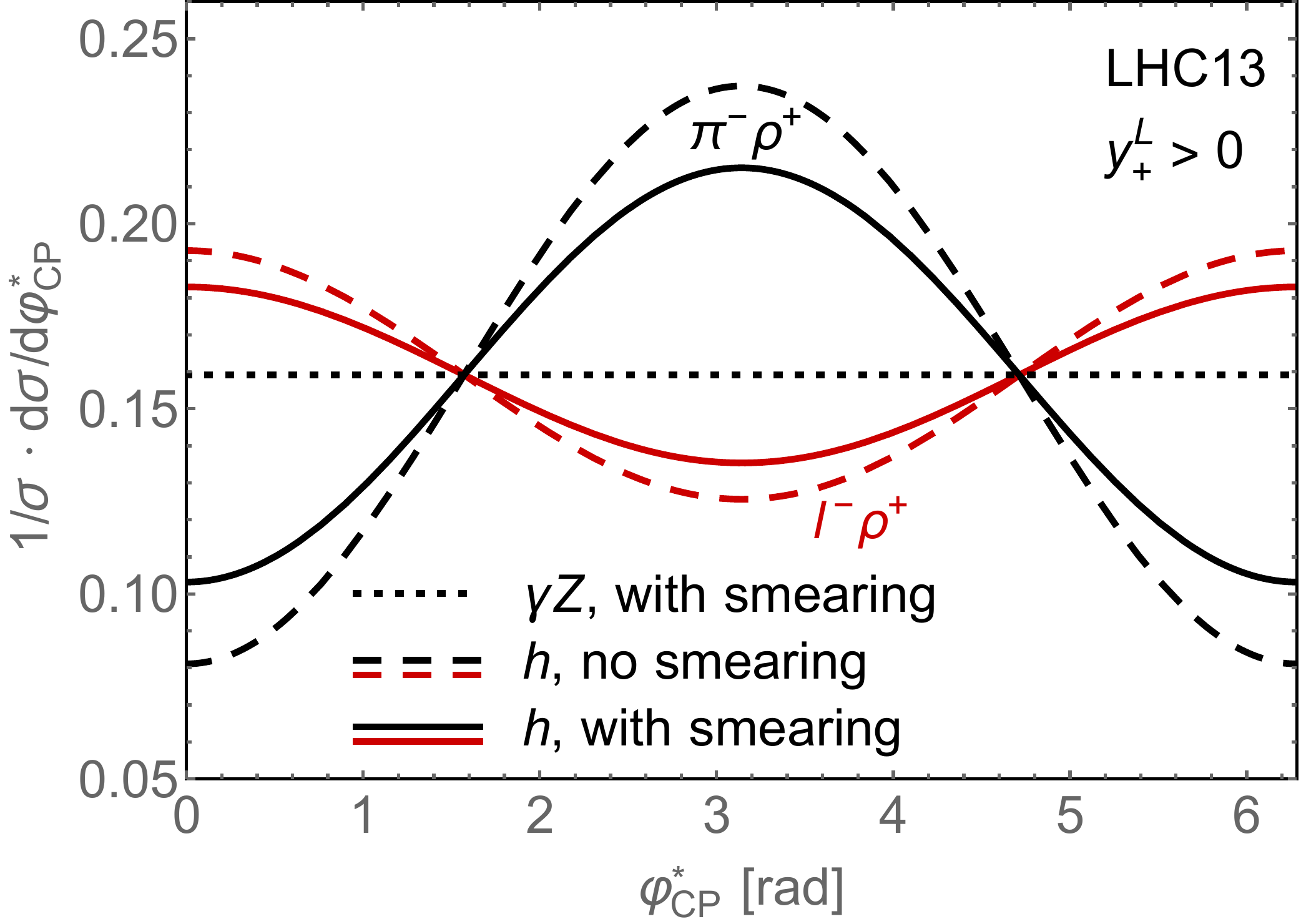}
\protect\caption{Impact of measurement uncertainties on $\varphi_{CP}^{*}$ distributions.
Left: CP-even Higgs boson $h \to \tau^{-}\tau^{+}\to\pi^{-}\rho^{+}$.
 Right: signal 
(solid black and red lines) and background 
(black dotted line) distributions for $\pi^{-}\rho^{+}$ and $l^{-}\rho^{+}$ final states.
\label{fig:h_impactrho_smearing} }
\end{figure}

 We have also investigated the impact of measurement uncertainties
on the $\varphi_{CP}^{*}$ distributions, both for the signal reactions
 and the Drell-Yan background. In Fig.~\ref{fig:h_impactrho_smearing}
  we show this impact for a CP-even Higgs boson and for the channels
  $\tau^{-}\tau^{+}\to\pi^{-}\rho^{+}$ and $\tau^{-}\tau^{+}\to l^{-}\rho^{+}$.
 We applied the cuts $p_{T}^{\rho^+}\ge20$~GeV, 
    $|\eta_{\rho^{+}}|, |\eta_{\pi^{+}}|\le2.5$  for $\tau^{+}\to\rho^{+}$
  and $q_{T}^{\pi^-,l^-}\ge 20$~GeV, $|\eta_{\pi^-}|, |\eta_{l^-}|\le2.5$ for
   $\tau^-\to \pi^-, l^-$.
   The experimental uncertainties are simulated with  a Gaussian smearing
of the impact parameter vectors of the charged tracks and the 4-momenta of the final 
 electrons, muons, charged and neutral pions as described in~\cite{Berge:2014sra}.
  For those $\tau$ decays where the impact
parameter method  is used to define $\varphi_{CP}^{*}$, 
 the primary vertex~(PV) is smeared 
     using  $\sigma_{z}^{PV}=20\mu$m, $\sigma_{tr}^{PV}=10\mu$m,
$\sigma_{tr}^{a^\pm}=10\mu$m, and the uncertainty on the intersection of the impact parameter vector and the charged
tracks is simulated with  $\sigma_{tr}^{\pi^-,l^-}=10\mu$m.
 The charged pion and lepton momenta are smeared\footnote{We consider a cone with opening angle $\theta$ around
the particle track and $\sigma_{\theta}$ denotes the smearing parameter around the track.}   using $\sigma_{\theta}^{a^\pm}=1$~mrad
and $\Delta E^{a^\pm}/E^{a^\pm}=5\%$.
 For the pions from $\tau^{\pm}\to\rho^{\pm}$ decay we
 use for the charged pion momenta $\sigma_{\theta}^{\pi\pm}=1$~mrad,
$\Delta E^{\pi^\pm}/E^{\pi^\pm}=5\%$ and for the neutral pion momenta
$\sigma_{\theta}^{\pi^{0}}=0.025/\sqrt{12}$~rad \cite{ATLAS-S08003},
 and $\Delta E^{\pi^{0}}/E^{\pi^{0}}=10\%$. 

As Fig.~\ref{fig:h_impactrho_smearing}, left, shows, the impact of this smearing 
  on the $\varphi_{CP}^{*}$ distribution for the $\tau^-\tau^+\to \pi^{-}\rho^{+}$ decay mode
  is rather small  if only the uncertainty
  of the $\pi^{0}$ momentum is taken into account (dotted black line). The uncertainty
   is dominated by the angular resolution
$\sigma_{\theta}^{\pi^{0}}$. The uncertainty of the $\pi^{\pm}$ momenta  has a much larger effect on
 the distribution (dashed black line) and is dominated by the smearing
of the $\pi^{-}$ impact parameter caused by the  uncertainties $\sigma_{z}^{PV}$,
$\sigma_{tr}^{PV}$, and $\sigma_{tr}^{\pi^\pm}$. 
  The solid black line shows the distribution taking into account all uncertainties.

In Fig.~\ref{fig:h_impactrho_smearing}, right, we display the effect
of the above smearing parameters on the distributions for the  decays to 
 $\pi^{-}\rho^{+}$ and $l^{-}\rho^{+}$, both for the signal and the Drell-Yan background.
 As shown in this figure the normalized $\varphi_{CP}^{*}$ distribution 
  of the $Z^*/\gamma^*$ background  
 is not affected by the smearing.
  This is in contrast to the case of the distributions for
   those $\tau^-\tau^+$ decay modes where for both $\tau$ decays the impact parameter
    method has to be used~\cite{Berge:2014sra}.

\section{Estimate of the expected precision on $\phi_{\tau}$}
\label{sec:phiestimate}
 In this section we estimate the precision with which the 
   mixing angle $\phi_{\tau}$ may be determined for the $h(125{\rm GeV})$ Higgs boson
     at the LHC (13 TeV) using the methods described in
     Sec.~\ref{susec:impactp}~-~\ref{susec:combi}.
 As in~\cite{Berge:2014sra}
we generate the $\varphi_{CP}^{*}$ distributions for Higgs-boson production
  by gluon-gluon fusion and for the  Drell-Yan background
$Z^*/\gamma^* \to\tau^{-}\tau^{+}$ for all major $\tau$-decay 
 modes \eqref{eq:taul} - \eqref{eq:taupi}.
  We use the $\rho$-decay plane method
 of Sec.~\ref{suec:rhodecpl} if both $\tau$ leptons decay to $\rho$ mesons and
 the combined method of Sec.~\ref{susec:combi} in case only one of the $\tau$ leptons
  decays to $\rho$. For all other $\tau$-decay modes
the impact parameter method is employed as described in Sec.~\ref{susec:impactp}. 

In order to compute the asymmetries  \eqref{phiCP_asym} for  the different decay channels,
 we apply the  following experimentally motivated cuts: 
 We require the $\tau$-pair invariant mass  $M_{\tau\tau}\ge100$~GeV for all $\tau^-\tau^+$
 decay channels. For the decays 
 $\tau^{\pm}\to\rho^{\pm}+\nu\to\pi^{\pm}+\pi^{0}+\nu$,
  we demand $p_{T}^{\rho^\pm}\ge20$~GeV and $|\eta_{\rho^\pm}|, |\eta_{\pi^{\pm}}|\le2.5$.
For the decays  $\tau^{\pm}\to l^{\pm}+2\nu$ and $\tau^{\pm}\to\pi^{\pm}+\nu$ 
 we apply the cuts $q_{T}^{\pi^\pm,l^\pm}\ge20$~GeV and 
$|\eta_{\pi^\pm}|, |\eta_{l^\pm}|\le2.5$. Furthermore, we assume that the longitudinal
 and transverse helicity states $a_{1}^{L,T\pm}$ of the $a_1$ resonance
  can be reconstructed and we use the cuts  $p_{T}^{a_{1}^{\lambda}}\ge20$~GeV and
   $|\eta_{a_{1}^{\lambda}}|\le2.5$ for the decays $\tau^{\pm}\to a_{1}^{L,T\pm}+\nu$.
 
   The experimental uncertainties are simulated by performing a Gaussian smearing
   of the 4-momenta and impact parameter vectors as described in Sec.~\ref{susec:combi}.

The asymmetries for those $h\to\tau^{+}\tau^{-}$ decay channels, where
at least one $\tau$ lepton decays to a $\rho$ meson,
 can be directly calculated from the smeared $\varphi_{CP}^{*}$ 
distributions    of the signal reactions, because 
   the respective  $\varphi_{CP}^{*}$ distribution of
 the Drell-Yan background  is flat, cf. Sec.~\ref{susec:combi}.
 For those channels, where the  impact
 parameter method has to be used both for the $\tau^-$ and $\tau^+$ 
  decay,  the shapes of the signal and background $\varphi_{CP}^{*}$ distribution 
  are deformed due to the smearing of the primary vertex. In these cases smeared asymmetries for
   the Higgs signal are computed  as described in~\cite{Berge:2014sra}.

\begin{table}
\begin{tabular}{|c|c|c|c|}
\hline 
$\tau\tau$ decay channel  & $\quad A_{S}\,\,[\%]\quad$  & $\quad\frac{S}{S+B}\quad$  & $\quad A_{S+B}\,\,[\%]\quad$ \tabularnewline
\hline 
\hline 
hadron-hadron  & $16.2$  & $0.5$  & $8.1$ \tabularnewline
\hline 
lepton-hadron  & $9.4$  & $0.5$  & $4.7$ \tabularnewline
\hline 
lepton-lepton  & $4.5$  & ${1}/{3}$  & $1.5$ \tabularnewline
\hline 
\end{tabular}\protect\caption{LHC13; Asymmetries for the hadron-hadron, lepton-hadron, and lepton-lepton
decay modes, obtained with the set of cuts and smearing parameters 
 described in the text. \label{tab:fin_Asymmetry_estimate} }
\end{table}

The resulting asymmetries of the signal distributions for the 
 hadron-hadron, lepton-hadron, and lepton-lepton final states
 are given in
the second column of Table~\ref{tab:fin_Asymmetry_estimate}. 
 The   third column of Table~\ref{tab:fin_Asymmetry_estimate}
 contains $S/(S+B)$ ratios taken from \cite{ATLtauconf} which we use here.
For the hadron-hadron and lepton-hadron decay channels we assume the ratio 
$S/B=1$ and $S+B=2\:{\rm events}/{\rm fb}$, while for  the lepton-lepton decay
modes we use $S/B=1/2$ and $S+B=2\:{\rm events}/{\rm fb}$. With these
numbers,  we obtain the asymmetries $A_{S+B}=A_S \times S/(S+B)$ given in column 4 of Table~\ref{tab:fin_Asymmetry_estimate}.

With $A_{S+B}$ and the expected total number of events 
we estimate with a procedure  described in~\cite{Berge:2014sra} the error $\Delta\phi_{\tau}$ with which 
 the mixing angle $\phi_{\tau}$ can be determined at the LHC.
 We obtain that $\phi_{\tau}$ can be measured with an uncertainty  of $15^{\circ}$
 ($9^{\circ}$) if an integrated luminosity of $150\, {\rm fb}^{-1}$ $(500\, {\rm fb}^{-1})$ 
  will be collected.  If eventually a luminosity
of $3\, {\rm ab}^{-1}$ could be achieved, the precision on $\phi_{\tau}$
may reach $3.6^{\circ}$.  The hadron-hadron decay modes of the $\tau^{+}\tau^{-}$
pair yield the highest precision: for instance, assuming an  
 integrated luminosity of $3\, {\rm ab}^{-1}$ we obtain $\Delta\phi_{\tau}\simeq 4^{\circ}$
  for these modes, while the hadron-lepton and lepton-lepton decay channels yield
  $\Delta\phi_{\tau}\simeq  7^{\circ}$ and  $\Delta\phi_{\tau}\simeq 22^{\circ}$, respectively.
   These results show that using both the impact parameter and the $\rho$-decay plane method
    and their combination improves the precision on  $\phi_{\tau}$ compared with 
     the achievable precision using only the impact parameter method.  With this method we obtained
      in~\cite{Berge:2014sra} for the combination of all decay channels   the estimates 
      $\Delta\phi_{\tau}\simeq 27^{\circ}$ ($150\, {\rm fb}^{-1}$), $\Delta\phi_{\tau}\simeq 14^{\circ}$ ($500\, {\rm fb}^{-1}$), 
       and $\Delta\phi_{\tau}\simeq 5^{\circ}$ ($3\, {\rm ab}^{-1}$).

 It should be noted that the above estimates depend on the Higgs-boson production
process and,  in particular, on the transverse momentum of the Higgs boson.
 For Higgs-boson events with large  large transverse  momenta, the asymmetries 
  decrease somewhat compared to the numbers given in Table~\ref{tab:fin_Asymmetry_estimate}. 
   In addition, the achievable
precision on $\phi_{\tau}$ strongly depends on the experimental resolution of the measurement 
of the impact parameters  and on the angular resolution of the  determination of the 
$\pi^{0}$ tracks.


\section{Impact on Two-Higgs-doublet models}

\label{sec:constrH} 
 In this section we analyze the impact of future measurements of the
  Higgs mixing angle $\phi_{\tau}$ and of the reduced $\tau$-Yukawa coupling
   strength $\kappa_{\tau}$
   on several SM extensions with a non-standard Higgs sector.

Neutral Higgs bosons with CP-violating couplings to quarks and leptons
appear in many SM extensions in a natural way. Here, we restrict ourselves
to non-supersymmetric SM extensions. (For recent discussions of Higgs-sector
CP violation in the context of supersymmetry, see \cite{Arbey:2014msa,Li:2015yla,King:2015oxa}.)

Two-Higgs-doublet models (2HDM) are among the simplest SM extensions
which allow for a reduced Yukawa coupling strength $\kappa_{\tau}\neq1$
and/or a mixing angle $\sin\phi_{\tau}\neq0$ in the interactions
of the 125 GeV Higgs resonance with $\tau$ leptons as parametrized
by \eqref{YukLa-phi}. These models are based on the SM gauge group
and the SM field content is extended by an additional Higgs doublet.
The physical particle spectrum of these models contains three neutral
Higgs particles $h_{i}$ $(i=1,2,3)$, one of which is to be identified
with the 125 GeV resonance, and a charged Higgs boson and its antiparticle,
$H^{\pm}$. CP-violating Yukawa couplings of neutral Higgs bosons
to quarks and leptons, in particular to $\tau$ leptons, appear in
these models in a natural way. (For a recent review of these models,
see, for instance, \cite{Branco:2011iw}.)

In the following we discuss the implications of future measurements
of the reduced Yukawa coupling strength $\kappa_{\tau}$ and of the
mixing angle $\phi_{\tau}$ on the parameter spaces of several variants
of 2HDM. As estimated in Sec.~\ref{sec:phiestimate}, we assume that
$\phi_{\tau}$ can be measured during the high-luminosity run of the
LHC (13 TeV) with a precision of $\Delta\phi_{\tau}=\pm 9^{\circ}$
and eventually
\footnote{This precision could also be achieved at a future high-luminosity
$e^{+}e^{-}$ collider, cf. \cite{Berge:2013jra}.%
} $\pm 4^{\circ}$.   For $\kappa_{\tau}$ a precision of $\pm4\,\%$
can be expected \cite{Dawson:2013bba}.

\begin{figure}[tb]
\noindent \centering{}\includegraphics[height=8cm]{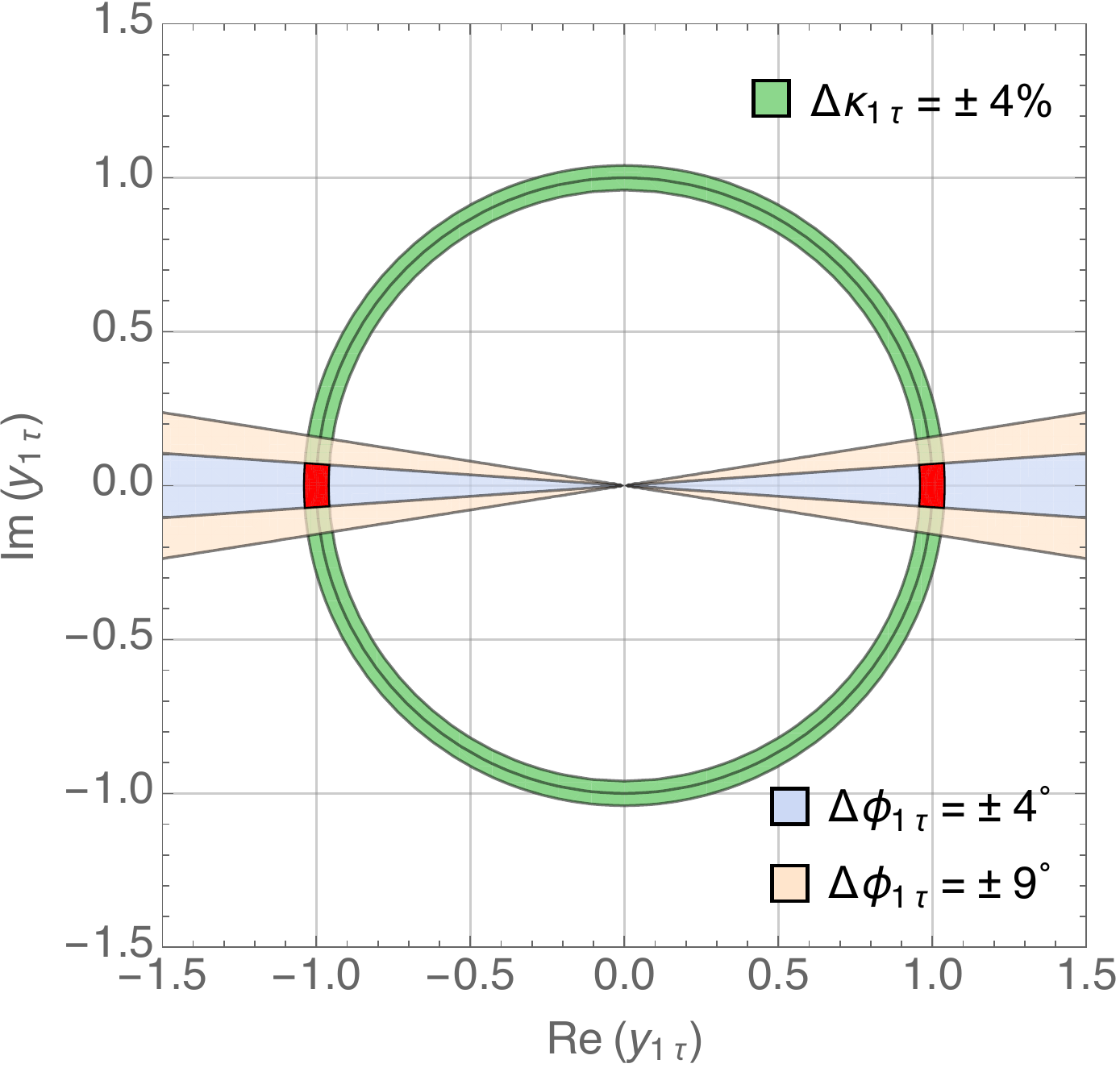}
\protect\protect\protect\caption{Allowed regions in the space of the reduced $\tau$-Yukawa couplings
${\rm Re}(y_{1\tau})$, ${\rm Im}(y_{1\tau})$ for assumed measurements
of $\kappa_{1\tau}=1.0\pm0.04$ and $\phi_{1\tau}=0^{\circ}\pm 9^{\circ}$
(grey segments) and $\phi_{1\tau}=0^{\circ}\pm 4^{\circ}$ (red segments),
respectively.
\label{fig:Rey_Imtau_prospect_A2HDM} 
}
\end{figure}

Let us first translate these expected experimental precisions into
bounds on the reduced Yukawa couplings to $\tau^{-}\tau^{+}$ of the
125 GeV resonance. In the following, we denote the 125 GeV Higgs boson
by $h_{1}$. Additional neutral Higgs bosons may exist -- we assume
here and in the following that they are non-degenerate with $h_{1}$.
The flavor-conserving Yukawa interactions of neutral Higgs bosons
$h_{i}$ to quarks and leptons $f=q,l$ may be parametrized in a model-independent
way as follows: %
\begin{equation}
{\cal L}_{Y}=-\frac{m_{f}}{\rm v}\left({\rm Re}(y_{if})~\bar{f}f+{\rm Im}(y_{if})~\bar{f}i\gamma_{5}f\right)h_{i}\,,\label{YukLa-phi-1}
\end{equation}
where a sum over $f$ and $i$ is understood. We concentrate here
on the reduced $\tau$ Yukawa couplings ${\rm Re}(y_{1\tau})$ and
${\rm Im}(y_{1\tau})$. Assuming that future measurements yield%
\footnote{In this section we put an additional label on $\kappa_{\tau}$ and
$\phi_{\tau}$ referring to the Higgs boson $h_{i}$.%
} $\kappa_{1\tau}=1.0\pm0.04$ and $\phi_{1\tau}=0^{\circ}\pm 9^{\circ}$,
respectively $\phi_{1\tau}=0^{\circ}\pm4 ^{\circ}$, one gets the very
small areas displayed in Fig.~\ref{fig:Rey_Imtau_prospect_A2HDM}
within which ${\rm Re}(y_{1\tau})$ and ${\rm Im}(y_{1\tau})$ must
lie. Notice that only the relative sign ${\rm Re}(y_{1\tau})$ and
${\rm Im}(y_{1\tau})$ is fixed, because the measured value of $\phi_{1\tau}$
cannot be distinguished from $\phi_{1\tau}+\pi$.

\subsection{The aligned 2HDM}

\label{suse:a2hdm} 

Phenomenologically viable 2HDM are usually constructed such that flavor-changing
neutral current (FCNC) interactions are absent at tree level. This
may be achieved by requiring Yukawa couplings such that none of the
right-chiral quark and lepton fields $f_{R}$ couples to both Higgs
doublets $\Phi_{1}$, $\Phi_{2}$. It can be enforced by assuming
an appropriately chosen discrete $Z_{2}$ symmetry (which is exactly
obeyed by the Yukawa Lagrangian), and there are several possible implementations
of such a symmetry (cf., for instance, \cite{Branco:2011iw}). (Below
we shall call these models `conventional 2HDM'.) Tree-level flavor
conservation of the neutral Higgs interactions can also be enforced
by allowing both Higgs doublets to couple to $f_{R}$ but assuming
that the Yukawa coupling matrices of $\Phi_{1}$ and $\Phi_{2}$ are
aligned in flavor space. While this is also an ad hoc assumption,
it is attractive from the phenomenological point of view: the resulting
model, the so-called aligned two-Higgs doublet model (A2HDM) formulated
in \cite{Pich:2009sp} contains as special cases all known 2HDM with
tree-level neutral-current flavor conservation, and it contains possible
new sources of CP violation. Apart from CP-violating mixing of the
neutral Higgs-boson states caused by a Higgs potential which is CP-violating already at tree level,
each of the aligned Yukawa matrices of the $u$-, $d$-type quark
and charged lepton sector may contain a CP phase which affects the
Yukawa couplings of the neutral Higgs bosons and those of the charged
Higgs boson.

The general gauge-invariant, hermitean, and renormalizable Higgs potential
$V(\Phi_{1},\Phi_{2})$ of 2HDM, which applies also to the A2HDM,
breaks the CP symmetry if no restriction on the parameters of $V$
is imposed. CP violation by $V$ is caused by complex couplings of
soft and hard $Z_{2}$-symmetry breaking terms in $V$. If this is
the case, the physical CP-even and -odd neutral Higgs fields mix and
the neutral fields respectively states $h_{1},h_{2},h_{3}$ in the
mass basis are CP mixtures already at tree level. In the context of the A2HDM we use, as
Ref.~\cite{Pich:2009sp}, the Higgs doublets $\Phi_{1}$, $\Phi_{2}$
in the so-called Higgs basis, where only $\Phi_{1}$ has a non-zero
vacuum expectation value. In this basis, the doublets can be brought
into the form 
\begin{equation}
\Phi_{1}=\left(G^{+},({\rm v} +S_{1}+iG^{0})/\sqrt{2}\right)^{T}\,,
 \qquad\Phi_{2}=\left(H^{+},(S_{2}+iS_{3})/\sqrt{2}\right)^{T}\,,\label{eq:higgsbas}
\end{equation}
where $G^{+}$ and $G^{0}$ are the Goldstone fields, $H^{+}$ is
the physical charged Higgs-boson field, and $S_{1}$, $S_{2}$ and $S_{3}$
are the physical neutral fields, which are CP-even and -odd, respectively.
They are related to the fields $h_{1},h_{2},h_{3}$ in the mass basis
by 
\begin{equation}
(h_{1},h_{2},h_{3})^{T}=R(S_{1},S_{2},S_{3})^{T}\,,\label{eq:hRS}
\end{equation}
where $R$ is a real orthogonal $3\times3$ matrix. The matrix elements
of $R$ depend on the parameters of the potential. Using \eqref{eq:hRS},
the Yukawa interactions \eqref{YukLa-phi-1} of the $h_{i}$ to quarks
and leptons, i.e., the reduced Yukawa couplings to $d$-type quarks,
charged leptons $l$, and $u$-type quarks are given by \cite{Pich:2009sp}:
\begin{eqnarray}
y_{id,l} & = & R_{i1}+(R_{i2}+iR_{i3})\zeta_{d,l}\,,\label{eq:yil}\\
y_{iu} & = & R_{i1}+(R_{i2}-iR_{i3})\zeta_{u}^{*}\,,\label{eq:yiu}
\end{eqnarray}
where the complex parameters $\zeta_{u},\zeta_{d},\zeta_{l}$ appear
in the alignment ansatz for the Yukawa matrices \cite{Pich:2009sp}
and provide additional sources of CP violation. The Yukawa interactions
of the charged Higgs bosons $H^{\pm}$, which involve also these complex
parameters and the Cabibbo-Kobayashi-Maskawa quark mixing matrix,
are not needed in the following.

The general reduced Yukawa couplings \eqref{eq:yil}, \eqref{eq:yiu}
contain a number of unknown parameters: three mixing angles which
parametrize $R$ and the three complex parameters $\zeta_{u},\zeta_{d},\zeta_{l}$.
We may identify the 125 GeV Higgs resonance with $h_{1}$ (see below).
The allowed regions of the reduced Yukawa couplings of $h_{1}$ to
$\tau$ leptons, ${\rm Re}(y_{1\tau}^{i})$ and ${\rm Im}(y_{1\tau})$,
which depend on five unknown parameters, are those displayed in Fig.~\ref{fig:Rey_Imtau_prospect_A2HDM}
if the envisaged experimental precisions can be attained. The determination
of the unknown parameters which determine the Yukawa couplings \eqref{eq:yil},
\eqref{eq:yiu} of the A2HDM requires a global fit, with the (future)
measurements of $\kappa_{1 \tau}$ and $\phi_{1\tau}$ being part of
the input, which is beyond the scope of this paper. We consider here
two special cases which are also of relevance for our purpose, namely the
A2HDM with complex parameter $\zeta_{l}$ and a Higgs potential that is CP-conserving at tree-level 
and the model with Higgs-sector CP violation and real $\zeta_{l}$.
 We discuss constraints on the parameters of these models resulting
solely from measurements of $\kappa_{1\tau}$ and $\phi_{1\tau}$.

\subsubsection{Tree-level CP-conserving Higgs potential and complex $\zeta_{l}$:}

In this case there is no mixing of $S_{3}$ with $S_{1}$, $S_{2}$
at tree level. The $3\times3$ matrix $R$ is now block-diagonal,
with $R_{13}=R_{23}=R_{31}=R_{32}=0$ and $R_{33}=1$. The mass eigenstates
$h_{1}$ and $h_{2}$, which result from the mixing of $S_{1}$ and
$S_{2}$, are CP-even and $h_{3}=S_{3}$ is CP-odd. (The assignment
of the CP quantum numbers to these states is determined by their tree-level
couplings to the weak gauge bosons.) Because the LHC results \cite{Khachatryan:2014jba,Aad:2015gba}
exclude that the 125 GeV Higgs resonance is a pseudoscalar (which
has no tree-level couplings to $W^{+}W^{-}$ and $ZZ$), it cannot
be identified with $h_{3}$. By convention we may identify it with
$h_{1}$. The reduced $\tau$-Yukawa couplings are in this case: 
\begin{eqnarray}
{\rm Re}(y_{i\tau})=R_{i1}+R_{i2}~{\rm Re}(\zeta_{l})\,, & \quad{\rm Im}(y_{i\tau})=R_{i2}~{\rm Im}(\zeta_{l})\,,\quad i=1,2\,,\label{eq:ComplYuk_ConservPot_2}\\
{\rm Re}(y_{3\tau})=-{\rm Im}(\zeta_{l})\,, & \quad{\rm Im}(y_{3\tau})={\rm Re}(\zeta_{l})\,.\label{eq:ComplYuk_ConservPot_4}
\end{eqnarray}
These equations show that the Higgs bosons $h_{i}$ can couple already at tree level  to
both scalar and pseudoscalar $\tau$ lepton currents, due to the additional
CP violation provided by the complex parameter $\zeta_{l}$ from the
aligned Yukawa sector, although the $h_{i}$ are CP eigenstates at tree level with
respect to their interactions with weak gauge bosons. Thus, also in this
special case of the A2DHM a nonzero value of $\sin\phi_{1\tau}$ is
possible. One should however notice that the CP-violating Yukawa couplings
 \eqref{eq:ComplYuk_ConservPot_2}, \eqref{eq:ComplYuk_ConservPot_4} induce at the 1-loop level
   CP-violating terms in the effective  Higgs potential which in turn lead to CP-violating
    mixing of the $h_{i}$ at the loop level. Therefore, beyond the tree level, the $h_i$ are no longer
     CP eigenstates.

The $2\times2$ orthogonal submatrix $(R_{ij})$ $(i,j=1,2)$ depends
on one parameter; i.e., for fixed~$i$, the two equations \eqref{eq:ComplYuk_ConservPot_2}
depend on three unknowns. Thus, in order to determine these parameters,
further experimental input is needed, apart from the measurement of
the reduced coupling strength $\kappa_{1\tau}$ and of $\phi_{1\tau}$.
Suppose this input implies that $R_{12}\neq0$. From Eq.~\eqref{eq:ComplYuk_ConservPot_2}
one gets for $i=1$, using the orthogonality of $R$ and $\left|{\rm Im}(y_{1\tau})/{\rm Im}(\zeta_{l})\right|\leq1$:
\begin{eqnarray}
0 & = & \kappa_{1\tau}\left({\rm Im}(\zeta_{l})\cos\phi_{1\tau}-{\rm Re}(\zeta_{l})\sin\phi_{1\tau}\right)\pm{\rm Im}(\zeta_{l})\sqrt{1-\left(\kappa_{1\tau}\sin\phi_{1\tau}/{\rm Im}(\zeta_{l})\right)^{2}}\,.\label{eq:reImzet}
\end{eqnarray}
 If it would turn out that  $\kappa_{1\tau}=1$, which corresponds to ${\rm Re}(y_{1\tau})=1$
and ${\rm Im}(y_{1\tau})=0$, Eq.~\eqref{eq:reImzet} is fulfilled
for all $\zeta_{l}$. In this case one could restrict the parameters
${\rm Re}(\zeta_{l}),{\rm Im}(\zeta_{l})$ using Eq.~\eqref{eq:ComplYuk_ConservPot_2}
if $R_{12}$ is known from some other measurement, and if it is non-zero
which is likely.

\begin{figure}[H]
\noindent \centering{}\includegraphics[width=7.5cm]{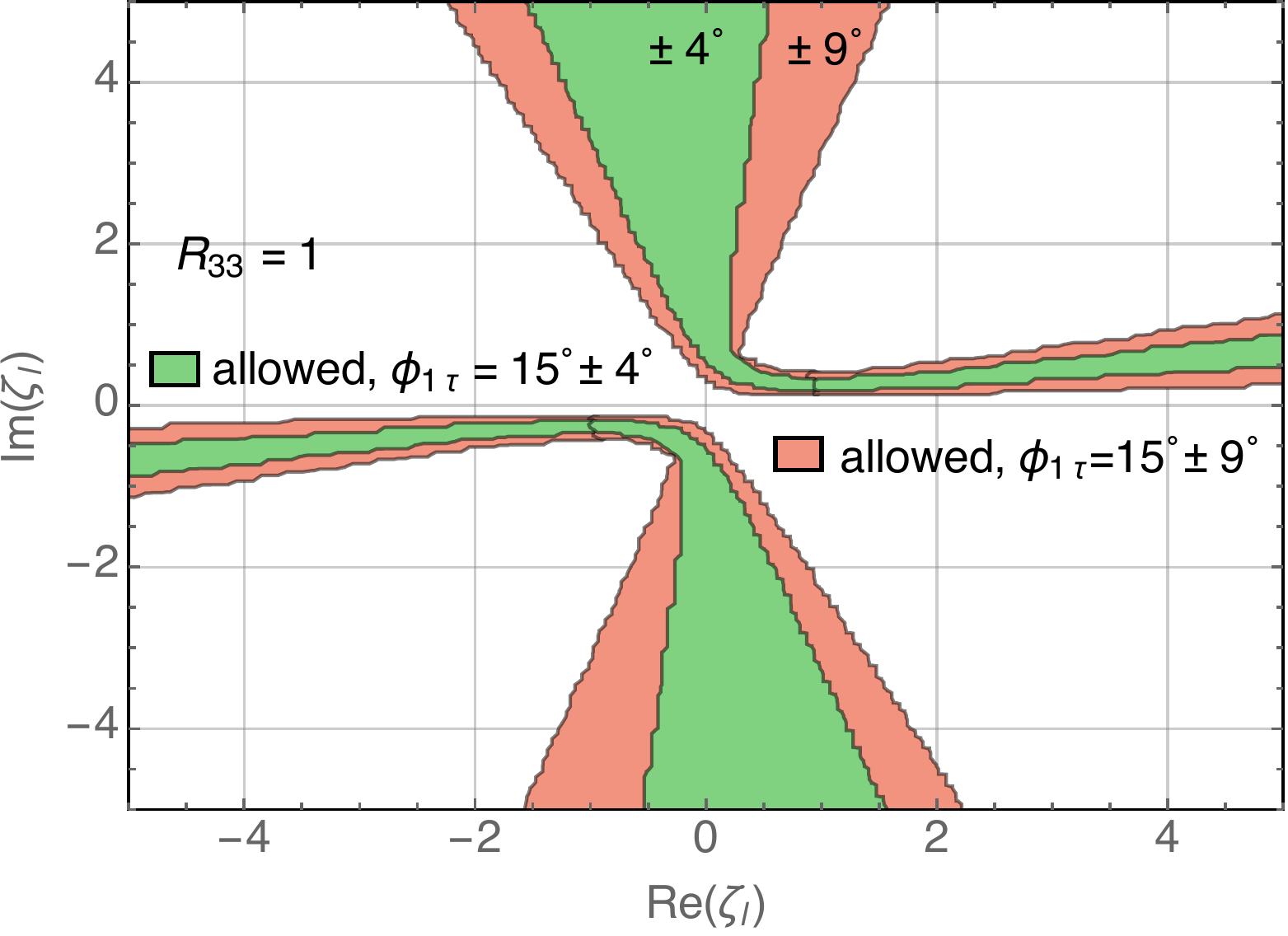}\protect\protect\caption{ Aligned 2HDM with CP-conserving Higgs potential but ${\rm Im}(\zeta_{l})\protect\ne0$.
For illustration we assume $\phi_{1\tau}$ is measured with a central
value of $\phi_{1\tau}=15^{\circ}$. The plot shows the resulting
allowed areas to which the parameters ${\rm Re}(\zeta_{l}),{\rm Im}(\zeta_{l})$are
restricted if $\Delta\phi_{1\tau}=\pm 9^{\circ}$ (red areas) and
$\Delta\phi_{1\tau}=\pm 4^{\circ}$ (green areas). 
\label{fig:Complex_Yuk_Pot_CPerhaltend} }
\end{figure}

However, if a non-zero value of $\phi_{1\tau}$ , e.g. $\phi_{1\tau}=15^{\circ}\pm 4^{\circ}$
will be measured, the allowed parameter range of ${\rm Re}(\zeta_{l})$
and ${\rm Im}(\zeta_{l})$ can be restricted without knowing $R_{12}$,
see Fig.~\ref{fig:Complex_Yuk_Pot_CPerhaltend}. If $\phi_{1\tau}$
can be measured with a precision of $9^{\circ}$ ($4^{\circ}$),
the red (green) areas in Fig.~\ref{fig:Complex_Yuk_Pot_CPerhaltend}
display the ranges in which the parameters ${\rm Re}(\zeta_{l})$
and ${\rm Im}(\zeta_{l})$ must then lie. Fig.~\ref{fig:Complex_Yuk_Pot_CPerhaltend}
is symmetric under a reflection at $\{{\rm Re}(\zeta_{l}),{\rm Im}(\zeta_{l})\}=\{0,0\}$
which corresponds to the sign choice in $R_{11}=\pm\sqrt{1-R_{12}^{2}}$.
Because $\phi_{1\tau}$ can be measured only modulo $\pi$, only the
relative sign of ${\rm Re}(\zeta_{l})$ and ${\rm Im}(\zeta_{l})$ is fixed. \\
 Existing experimental upper bounds on the electric dipole moments
of the neutron and of atoms/molecules provide upper bounds on $|{\rm Im}(\zeta_{u}^{*}\zeta_{d})|$
and $|{\rm Im}(\zeta_{q}^{*}\zeta_{l}|$, which depend, however,
on the masses of the neutral and charged Higgs bosons \cite{Jung:2013hka}.

\subsubsection{CP-violating Higgs potential and real $\zeta_{l}$:}

Another limiting case of the A2HDM, which is also relevant for CP
violation in the $\tau$ decays of the 125 GeV resonance, is the model
with CP-violating tree-level Higgs potential but real alignment parameters $\zeta_{u},\zeta_{d},\zeta_{l}$.
In this case the reduced Yukawa couplings of the $h_{i}$ read: 
\begin{eqnarray}
{\rm Re}(y_{id,l})=R_{i1}+R_{i2}\zeta_{d,l}\,, & \qquad{\rm Im}(y_{id,l})=R_{i3}\zeta_{d,l}\,,\label{eq:rezl1}\\
{\rm Re}(y_{iu})=R_{i1}+R_{i2}\zeta_{u}\,, & \qquad{\rm Im}(y_{iu})=-R_{i3}\zeta_{u}\,.\label{eq:rezeu1}
\end{eqnarray}
The couplings \eqref{eq:rezl1} and \eqref{eq:rezeu1} depend on six
real parameters. We assume that $\kappa_{1\tau}$ and $\phi_{1\tau}$
will be measured with the precision as stated in the preceding subsection.
For $\zeta_{l}\ne0$, Eqs.~\eqref{eq:rezl1} imply
\begin{eqnarray}
0\,\,= & \,\,\,\kappa_{1\tau}\left(R_{12}\sin\phi_{1\tau}-R_{13}\cos\phi_{1\tau}\right)\,\,\pm & \,\, R_{13}\sqrt{1-R_{12}^{2}-R_{13}^{2}}\,.\label{eq:relR123}
\end{eqnarray}
\begin{figure}[H]
\includegraphics[width=7.5cm]{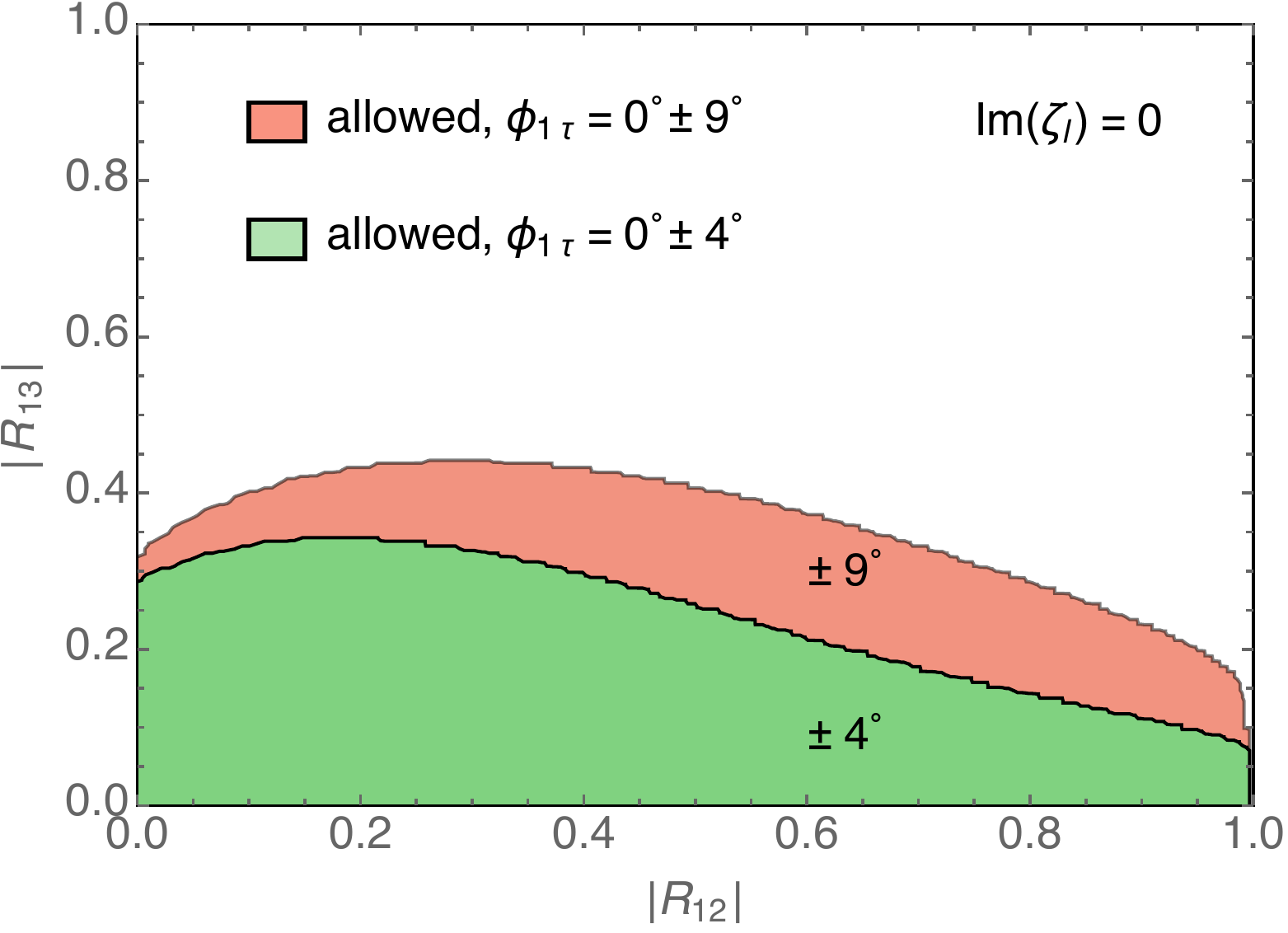}\hspace{0.5cm}\includegraphics[width=7.5cm]{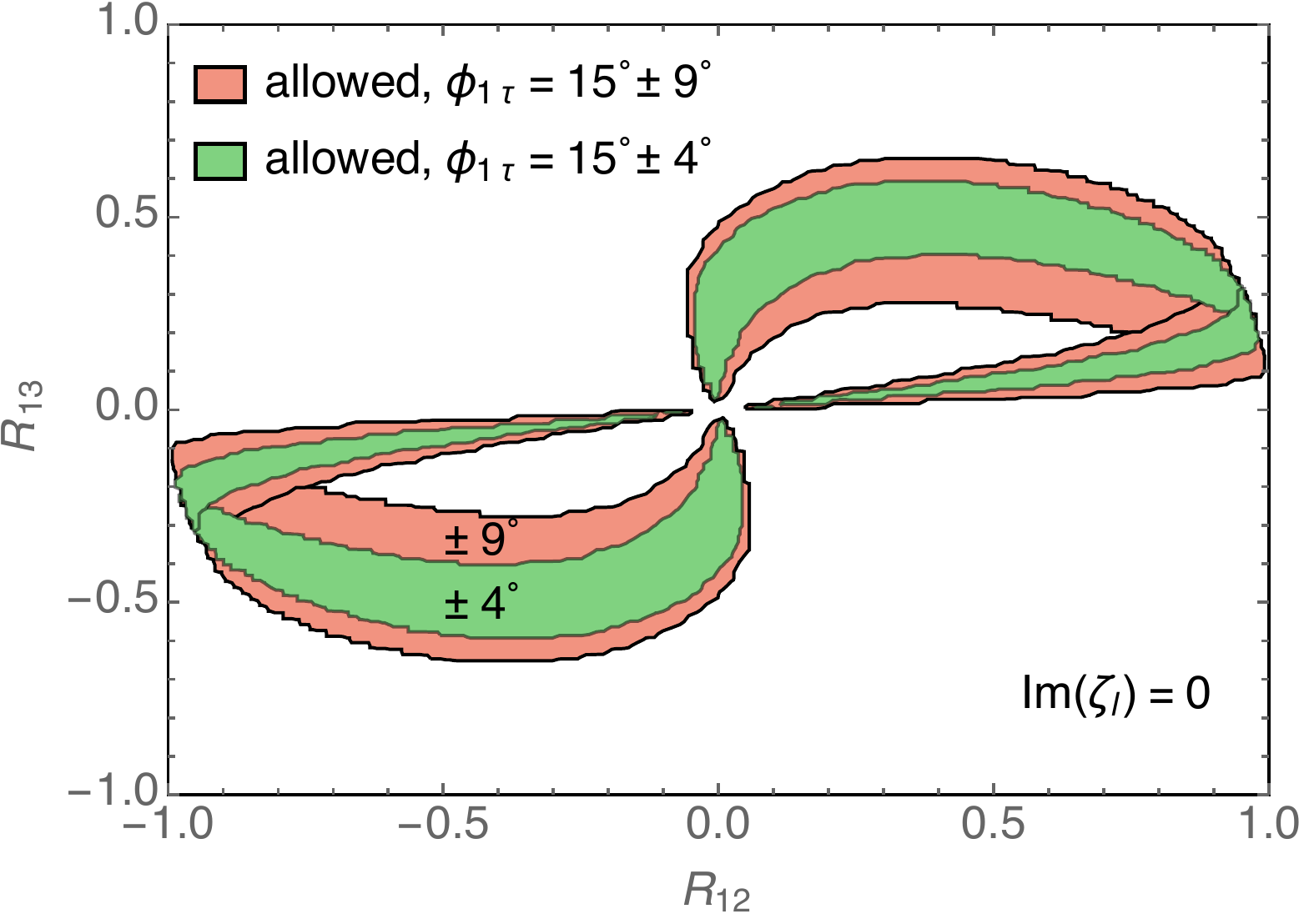}
\protect\protect\protect\caption{A2HDM with CP-violating Higgs potential and ${\rm Im}(\zeta_{l})=0$.
Constraints on the mixing matrix elements $R_{12}$ and $R_{13}$
assuming that measurements yield $\phi_{1\tau}=0^{\circ}$ (left plot)
or $\phi_{1\tau}=15^{\circ}$ (right plot). The red (green) areas
show the allowed regions which would remain if $\phi_{1\tau}$ will
be measured with a precision of $\Delta\phi_{1\tau}=\pm 9 ^{\circ}$
($\Delta\phi_{1\tau}=\pm 4^{\circ}$). Notice that in the right plot
the allowed regions are restricted to $|R_{12}|<1$. 
\label{fig:Imzeta0_Ri1Ri3_Ri2Ri3_A2HDM} }
\end{figure}

Relation \eqref{eq:relR123} leads to constraints on the Higgs mixing
matrix elements $R_{12}$ and $R_{13}$ which are illustrated in the
left and right plot in Fig.~\ref{fig:Imzeta0_Ri1Ri3_Ri2Ri3_A2HDM}.
We recall that $\phi_{1\tau}$ will be experimentally determined with
our method only modulo $\pi$. Thus, if the central value of $\phi_{1\tau}$
turns out to be zero, only the moduli $R_{12}$ and $R_{13}$ are
constrained by \eqref{eq:relR123}. If $\phi_{1\tau}$ turns out to
be non-zero, the relative sign of $R_{12}$ and $R_{13}$ is fixed
by \eqref{eq:relR123}.

The mixing matrix elements $R_{12}$ and $R_{13}$ are restricted
already by measurements of $CP$-even observables \cite{Celis:2013ixa}
and by existing upper bounds on electric dipole of the neutron and
of atoms/molecules \cite{Jung:2013hka}. These constraints in the
$R_{12}-R_{13}$ plane are different from those illustrated in Fig.~\ref{fig:Imzeta0_Ri1Ri3_Ri2Ri3_A2HDM}.
Thus, the measurement of $\phi_{1\tau}$ would, if interpreted within
this model, either be evidence for Higgs-sector CP violation or further
constrain this scenario.


\subsection{Conventional 2HDM with neutral flavor conservation\label{sub:Conventional-2HDM-with}}


As shown in \cite{Pich:2009sp} the aligned 2HDM contains as special
cases the known `conventional' 2HDM with tree-level neutral flavor
conservation based on (approximate) $Z_{2}$ symmetries. We briefly
discuss the implications of future measurements of $\kappa_{1\tau}$
and $\phi_{1\tau}$ on the parameters of these models.

\subsubsection{Type-I and type-II model:}

In the 2HDM of type-I, only the doublet $\phi_{2}$ is coupled to
fermions, while in the model of type-II, $\Phi_{1}$ is coupled to
$d_{R}$, $l_{R}$ and $\Phi_{2}$ is coupled to $u_{R}$. In these
models no additional CP violation besides the Kobayashi-Maskawa phase
arises from the Yukawa matrices. The parameters $\zeta_{u},\zeta_{d},\zeta_{l}$,
which are real in these models, are no longer independent, but are
given \cite{Pich:2009sp} in terms of the parameter $\beta=\arctan({\rm v}_{2}/{\rm v}_{1})$,
where ${\rm v}_{1}$ and ${\rm v}_{2}$ are the vacuum expectations of the neutral
components of the doublets $\Phi_{1}$ and $\Phi_{2}$, respectively.
Thus, CP-violating effects caused by the Yukawa couplings of the neutral
Higgs bosons $h_{i}$ require for both models (and also for other
conventional 2HDM) mixing of CP-even and-odd neutral states caused
by a CP-violating Higgs potential. In this context it is customary
to start from the usual representation of the doublets, 
 $\Phi_{j}=(\varphi_{j}^{+},({\rm v}_{j}+\varphi_{j}+i\chi_{j})/\sqrt{2})^{T}$,
$(j=1,2)$, and diagonalize the $3\times3$ squared mass matrix of
the physical neutral Higgs bosons in the basis $\varphi_{1},\varphi_{2},A,$
where $A=-\sin\beta\chi_{1}+\cos\beta\chi_{2}$. This diagonalization
is accomplished by an orthogonal matrix $O$. The neutral Higgs fields
$h_{i}$ in the mass basis are given by 
\begin{equation}
(h_{1},h_{2},h_{3})^{T}=O(\varphi_{1},\varphi_{2},A)^{T}\,.\label{eq:hOphi}
\end{equation}
The matrix $O$ is related to the matrix $R$ defined in \eqref{eq:hRS}
by 
\begin{equation}
O=R\left(\begin{array}{ccc}
\cos\beta & \sin\beta & 0\\
-\sin\beta & \cos\beta & 0\\
0 & 0 & 1
\end{array}\right)\,.\label{eq:relOR}
\end{equation}
For the type-I and type-II 2HDM with neutral Higgs sector CP violation
the reduced scalar and pseudoscalar Yukawa couplings \eqref{YukLa-phi-1}
of the neutral Higgs bosons $h_{i}$ are given
in terms of the matrix elements of $O$ as listed in Table~\ref{tab:YUko12},
 cf. \cite{Bernreuther:1992dz,Accomando:2006ga}.

\vspace{2mm}

\begin{table}[h]
\begin{centering}
\protect\protect\caption{Reduced Yukawa couplings \eqref{YukLa-phi-1} to quarks and leptons
of the neutral Higgs bosons $h_{i}$ in the type-I and type-II 2HDM.\label{tab:YUko12}}

\par\end{centering}

\centering{}\vspace{1mm}
\begin{tabular}{c|cccc}
 & ${\rm Re}(y_{iu})$  & ${\rm Re}(y_{id})={\rm Re}(y_{il})$  & ${\rm Im}(y_{iu})$  & ${\rm Im}(y_{id})={\rm Im}(y_{il})$ \tabularnewline
\hline 
\hline 
Type-I  & $O_{i2}/\sin\beta$  & $O_{i2}/\sin\beta$  & $-O_{i3}\cot\beta$  & $O_{i3}\cot\beta$ \tabularnewline
Type-II  & $O_{i2}/\sin\beta$  & $O_{i1}/\cos\beta$  & $-O_{i3}\cot\beta$  & $-O_{i3}\tan\beta$ \tabularnewline
\hline 
\end{tabular}
\end{table}

In the type-I and -II 2HDM the neutral Higgs Yukawa couplings depend
on 4 real parameters: three angles with which the matrix $O$ can
be parametrized, and $\tan\beta$. Other couplings of the $h_{i}$
and of $H^{\pm}$ also depend on (some of) these parameters. Constraints
from $B$ physics and $B_{0}-{\bar{B}}_{0}$ mixing imply that $\tan\beta$
should be larger than $0.5-0.7$ \cite{Eberhardt:2013uba}. Recently
a number of investigations were made within type-I and -II 2HDM with
neutral Higgs sector CP violation, including \cite{Barroso:2012wz,Cheung:2014oaa,Grzadkowski:2014ada,Fontes:2015mea,Inoue:2014nva,Fontes:2015xva,Chen:2015gaa},
on how LHC and B physics data and the present upper limits on the
electric dipole moments of the neutron \cite{Baker:2006ts} and of
the electron \cite{Baron:2013eja} constrain the mixing angles of
$O$. In the following we employ the parametrization of $O$ in terms
of three angles, $\alpha$, $\alpha_{c}$, and $\alpha_{b}$ in the
convention used in \cite{Inoue:2014nva,Chen:2015gaa}. In the CP-conserving
limit of the type-I and -II 2HDM $\alpha_{c}=\alpha_{b}=0$. The bounds
on the CP angles $\alpha_{c},\alpha_{b}$ derived in \cite{Inoue:2014nva,Chen:2015gaa}
depend, in particular, on the masses of the $h_{i}$ and $H^{+}$.
A measurement of the mixing angle $\phi_{1\tau}$ with some precision
in the $\tau$ decays of the 125 GeV Higgs resonance would yield significant
information on neutral Higgs-sector CP violation in the context of
these models -- independent of the masses of the charged and the other
neutral Higgs bosons.

\begin{figure}[t]
\includegraphics[height=5cm]{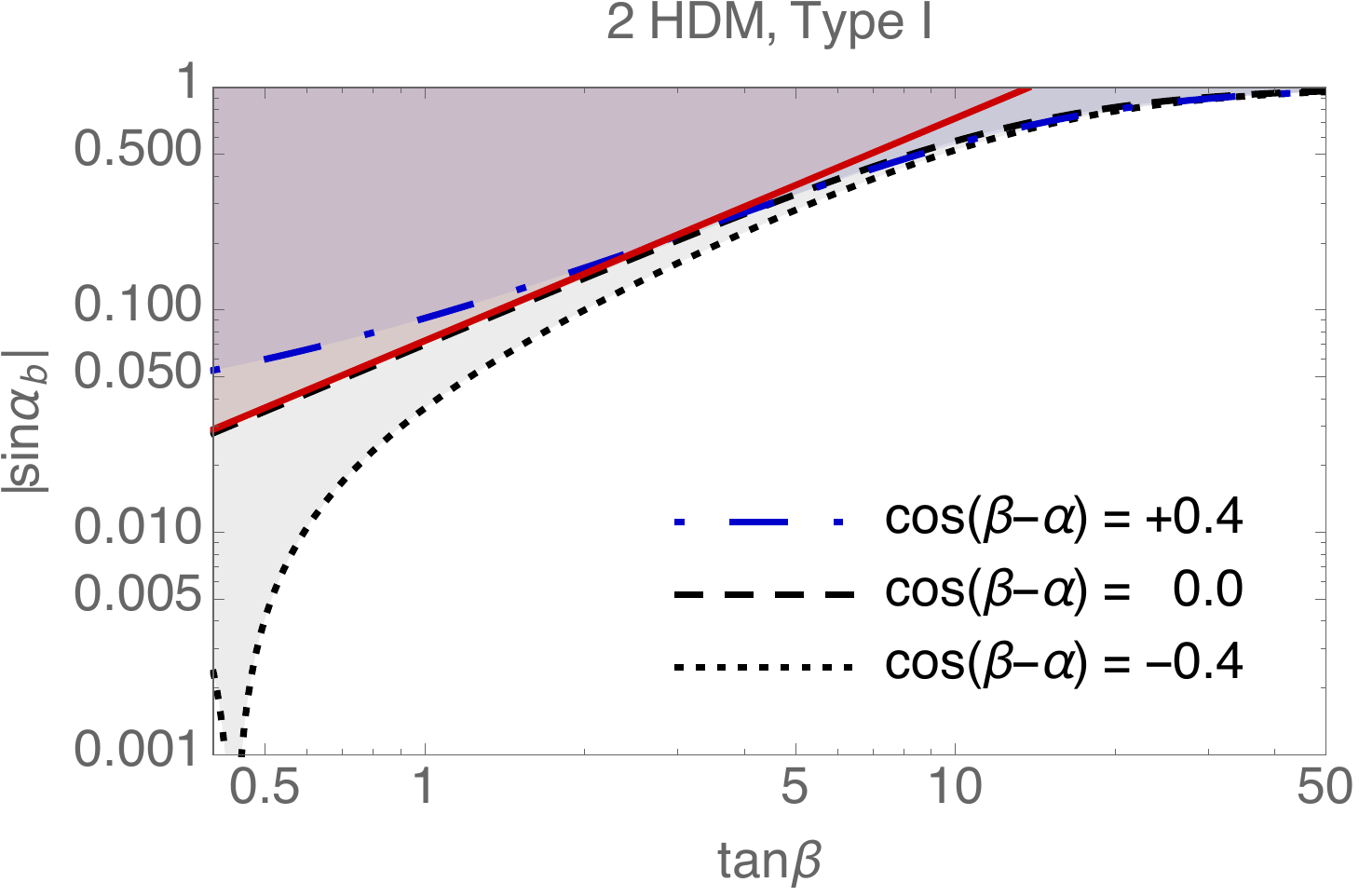}\hspace{0.3cm}\includegraphics[height=5cm]{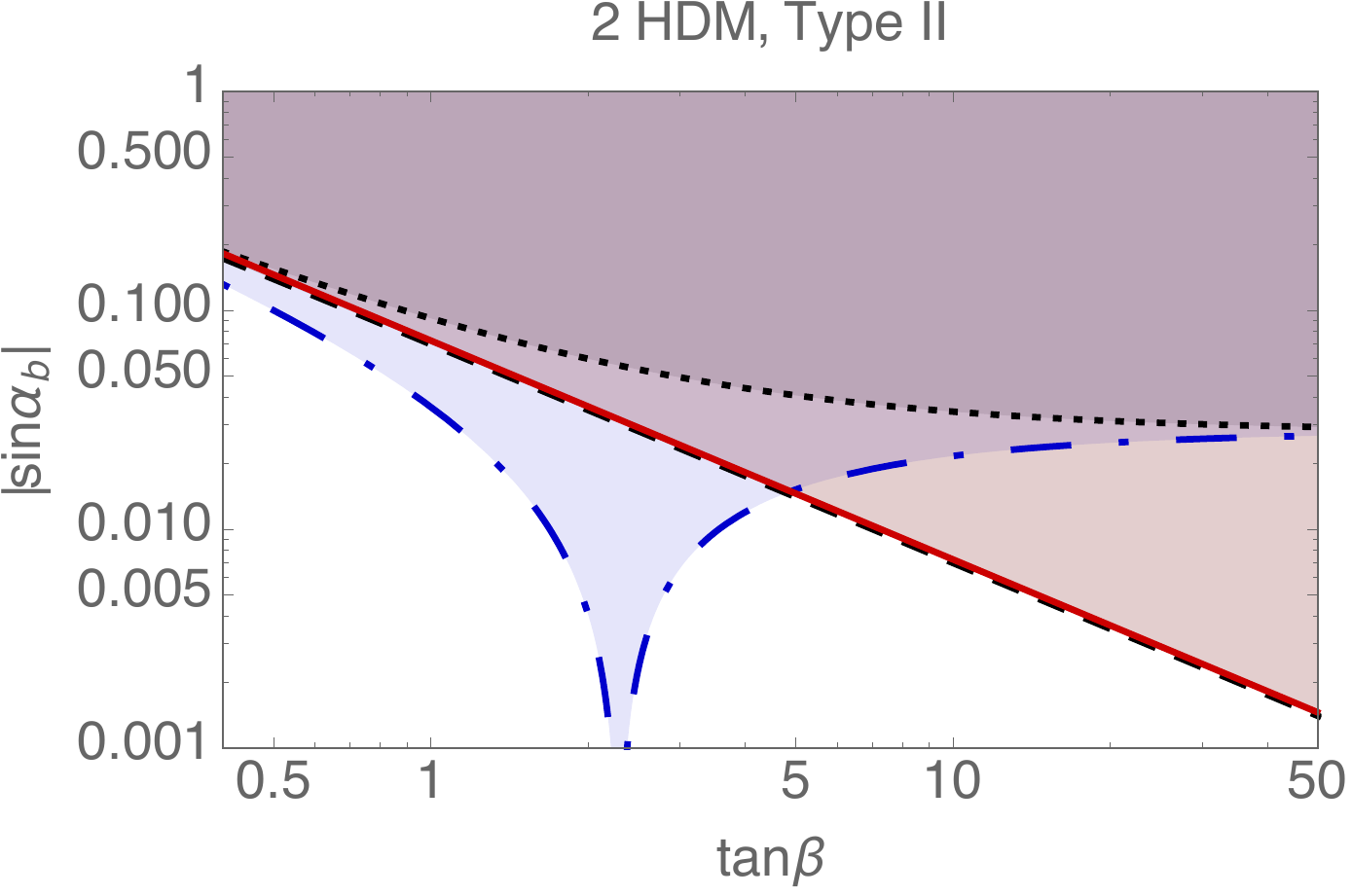}
\protect\protect\protect\caption{Exclusion ranges (colored) in the $|\sin\alpha_{b}|$, $\tan\beta$
parameter space, assuming that measurements yield $\phi_{1\tau}=0^{\circ}\pm 4^{\circ}$.
The area between the solid red lines and the top of the plots will
then be excluded, irrespective of the value of the mixing angle $\alpha$.
The dot-dashed blue, dashed black, and dotted black lines are the
boundaries of $\alpha$-dependent exclusion ranges for fixed values
of $\cos(\beta-\alpha)=0.4,0,$ and $-0.4$, respectively. Left: type
I model; right: type II model. 
\label{fig:Dawson_2DHM_Exclusion} }
\end{figure}

This can be seen as follows. We identify the 125 GeV Higgs with $h_{1}$.
(The following argumentation can also be applied to the other neutral
Higgs bosons.) For the Yukawa couplings of $h_{1}$ the matrix elements
$O_{1j}$ are relevant. Using the parametrization of $O$ as in \cite{Inoue:2014nva,Chen:2015gaa},
we have $O_{11}=-\sin\alpha\cos\alpha_{b}$, $O_{12}=\cos\alpha\cos\alpha_{b}$,
$O_{13}=\sin\alpha_{b}$. That is, these matrix elements do not depend
on $\alpha_{c}$. Using that $\tan\phi_{1\tau}={\rm Im}(y_{1l})/{\rm Re}(y_{1l})$
and $\kappa_{1,\tau}=[({\rm Re}(y_{1l}))^{2}+({\rm Im}(y_{1l}))^{2}]^{1/2}$,
and using the reduced Yukawa couplings of Table~\ref{tab:YUko12},
we get for\vspace{-4mm}
 
\begin{eqnarray}
{\rm Type\, I:} & \qquad & \tan\alpha_{b}=\frac{\cos\alpha}{\cos\beta}~\tan\phi_{1\tau}\,,\label{eq:phitau-I}\\
 &  & \sin\alpha_{b}=\kappa_{1\tau}\tan\beta~\sin\phi_{1\tau}\,,\label{eq:kapptau-I}\\
\nonumber \\
{\rm Type\, II:} & \qquad & \tan\alpha_{b}=\frac{\sin\alpha}{\sin\beta}~\tan\phi_{1\tau}\,,\label{eq:phitau-II}\\
 &  & \sin\alpha_{b}=-\kappa_{1\tau}\cot\beta~\sin\phi_{1\tau}\,.\label{eq:kapptau-II}
\end{eqnarray}
Let's assume that future measurements would yield $\phi_{1\tau}=0^{\circ}\pm 4^{\circ}$
and $\kappa_{1\tau}=1\pm0.04$. The resulting constraints on the two-dimensional
$\{|\sin\alpha_{b}|,\tan\beta\}$ parameter space of the type-I and
type-II 2HDM are shown in Fig.~\ref{fig:Dawson_2DHM_Exclusion},
left and right, respectively. The white (colored) areas in Fig.~\ref{fig:Dawson_2DHM_Exclusion},
which depend on the assumptions on $\cos(\beta-\alpha)$ stated in
the caption of the figure, are the remaining allowed (excluded) regions.
The area above the solid red line would be excluded, independent of
the values of the mixing angle $\alpha$.

Constraints on $\sin\alpha_{b}$ and $\tan\beta$ were derived in
\cite{Chen:2015gaa} using LHC results on $h_{1}$, on searches for
$h_{2,3}$, and the experimental upper bounds on the EDM of the neutron
and the electron. In some regions the constraints that will be obtainable solely
from the measurement of $\phi_{1\tau}$ and $\kappa_{1\tau}$ are
complementary to the constraints derived in \cite{Chen:2015gaa}.
In the region around $\tan\beta\sim1$ the determination of $\phi_{1\tau}$
would actually provide a stronger constraint. One should also recall
that the measurement of $\phi_{1\tau}$ probes Higgs-sector CP violation
directly, while EDM of leptons and hadrons can be induced also by
other non-standard CP-violating interactions.

\subsubsection{Flipped, lepton-specific, and inert model:}

The flipped 2HDM is defined by the coupling prescriptions $d_{R}\leftrightarrow\Phi_{1}$
and $u_{R},l_{R}\leftrightarrow\Phi_{2}$, while the lepton-specific
model is defined by $l_{R}\leftrightarrow\Phi_{1}$ and $u_{R},d_{R}\leftrightarrow\Phi_{2}.$
Overviews on the phenomenology of these models are given in \cite{Aoki:2009ha,Branco:2011iw}.
In the flipped 2HDM the Yukawa couplings of the $h_{i}$ to $u$-type
quarks and charged leptons are identical to the corresponding ones
of the type-I model, while the Yukawa couplings to $d$-type quarks
are those of type-II. In the lepton-specific model, the Yukawa couplings
of the $h_{i}$ to quarks are those of the type-I model, while the
Yukawa couplings to charged leptons are those of type-II. Therefore,
our discussion in the previous subsection of how future measurements
of $\kappa_{1\tau}$ and $\phi_{1\tau}$ will provide information
on the CP-violating Higgs-mixing parameter $|\sin\alpha_{b}|$ within
the type-I and type-II models can be taken over. The left plot and
right plot in Fig.~\ref{fig:Dawson_2DHM_Exclusion} applies also
to the flipped and lepton-specific 2HDM, respectively. In these models,
the branching ratio of $h_{1}\to\tau^{+}\tau^{-}$ can be larger than
that of $h_{1}\to b{\bar{b}}$, contrary to the case of type-I and
type-II models.

Finally, a remark on the so-called inert model, which is a 2HDM with
an unbroken $Z_{2}$ symmetry \cite{Branco:2011iw}. The Higgs-boson
spectrum of this model contains a neutral state which is stable and
will therefore contribute to the dark matter density. In this model,
the tree-level Higgs potential is CP-invariant by virtue of the imposed
$Z_{2}$ symmetry. Thus the inert 2HDM predicts $\sin\phi_{1\tau}=0$
in $h_{1}\to\tau^{+}\tau^{-}$.


\section{Conclusions}

\label{sec:conlusions} 

 We have investigated the precision with which the CP nature of the
 125 GeV Higgs boson $h$ can be determined in its decays to $\tau^-\tau^+$
  at the LHC (13~TeV).
  We have taken into account all major $\tau$-decays to charged prongs.
   Contrary to \cite{Berge:2014sra} where the impact parameter method was used for all
    $\tau$ decays in order to define an observable $\varphi^*_{CP}$ 
    which is sensitive to possible scalar-pseudoscalar Higgs mixing parametrized by
     an angle $\phi_\tau$, we combined in this paper two methods: We used the
      $\rho$-decay plane method  for $\tau^\mp\to\rho^\mp$ and the impact parameter method for all
       other major $\tau$ decays. This combination leads to an increase of the sensitivity of
   $\varphi^*_{CP}$ to the mixing angle $\phi_\tau$. In estimating this sensitivity we took into account
    the contributions from the Drell-Yan background, and we have analyzed by Monte-Carlo simulation how
     measurement uncertainties affect the signal and background $\varphi^*_{CP}$ distributions.
     We found that the mixing angle $\phi_\tau$ can be determined with an uncertainty of 
     $\Delta\phi_\tau\simeq 15^{\circ}$
 ($9^{\circ}$) at the LHC with an integrated luminosity of $150\, {\rm fb}^{-1}$ $(500\, {\rm fb}^{-1})$.
  At the high-luminosity LHC with an integrated luminosity of  $3\, {\rm ab}^{-1}$ 
  a precision of $\approx 4^{\circ}$ on $\phi_{\tau}$ could be reached.
     
  The precise measurement of the mixing angle $\phi_\tau$ is of great importance for possible physics beyond
   the Standard Model, because a non-zero value of $\phi_\tau$ would signify a new type of CP-violating interaction arising
   from an extended Higgs sector. We have analyzed the impact of future measurements of $\phi_\tau$ and of the $h\tau\tau$
   Yukawa coupling strength $\kappa_\tau$ on  the parameter spaces of
   a number of 2-Higgs-doublet extensions of the SM with Higgs-sector CP violation, namely the aligned 2HDM and several
   conventional 2 HDM with tree-level neutral flavor conservation. Contrary to the information, respectively the constraints 
    which arise from the measurements of the electric dipole moments of atoms/molecules and the neutron, the  measurement
    of $\phi_\tau$ yields direct information on CP-violating neutral Higgs-boson mixing, independent of the mass-values of the
     other Higgs bosons.

\section*{Acknowledgments}

We thank Philipp Bechtle, Klaus Desch, Maike Hansen, Peter Wagner
and the members of the $m_{\tau\tau}$ working group of the Helmholtz
alliance ``Physics at the Terascale'' for helpful discussions. S.
B. was supported by B.M.B.F. and S.K. by Deutsche Forschungsgemeinschaft
through Graduiertenkolleg GRK 1675.

\bibliographystyle{apsrev}
\bibliography{bibliography_BBK}

\begin{thebibliography}{53}
\expandafter\ifx\csname natexlab\endcsname\relax\def\natexlab#1{#1}\fi
\expandafter\ifx\csname bibnamefont\endcsname\relax
  \def\bibnamefont#1{#1}\fi
\expandafter\ifx\csname bibfnamefont\endcsname\relax
  \def\bibfnamefont#1{#1}\fi
\expandafter\ifx\csname citenamefont\endcsname\relax
  \def\citenamefont#1{#1}\fi
\expandafter\ifx\csname url\endcsname\relax
  \def\url#1{\texttt{#1}}\fi
\expandafter\ifx\csname urlprefix\endcsname\relax\def\urlprefix{URL }\fi
\providecommand{\bibinfo}[2]{#2}
\providecommand{\eprint}[2][]{\url{#2}}

\bibitem[{\citenamefont{Aad et~al.}(2012)}]{Aad:2012tfa}
\bibinfo{author}{\bibfnamefont{G.}~\bibnamefont{Aad}} \bibnamefont{et~al.}
  (\bibinfo{collaboration}{ATLAS Collaboration}), \bibinfo{journal}{Phys.Lett.}
  \textbf{\bibinfo{volume}{B716}}, \bibinfo{pages}{1} (\bibinfo{year}{2012}),
  \eprint{1207.7214}.

\bibitem[{\citenamefont{Chatrchyan et~al.}(2012)}]{Chatrchyan:2012ufa}
\bibinfo{author}{\bibfnamefont{S.}~\bibnamefont{Chatrchyan}}
  \bibnamefont{et~al.} (\bibinfo{collaboration}{CMS Collaboration}),
  \bibinfo{journal}{Phys.Lett.} \textbf{\bibinfo{volume}{B716}},
  \bibinfo{pages}{30} (\bibinfo{year}{2012}), \eprint{1207.7235}.

\bibitem[{\citenamefont{Khachatryan et~al.}(2015)}]{Khachatryan:2014jba}
\bibinfo{author}{\bibfnamefont{V.}~\bibnamefont{Khachatryan}}
  \bibnamefont{et~al.} (\bibinfo{collaboration}{CMS}), \bibinfo{journal}{Eur.
  Phys. J.} \textbf{\bibinfo{volume}{C75}}, \bibinfo{pages}{212}
  (\bibinfo{year}{2015}), \eprint{1412.8662}.

\bibitem[{\citenamefont{Aad et~al.}(2015)}]{Aad:2015gba}
\bibinfo{author}{\bibfnamefont{G.}~\bibnamefont{Aad}} \bibnamefont{et~al.}
  (\bibinfo{collaboration}{ATLAS}) (\bibinfo{year}{2015}), \eprint{1507.04548}.

\bibitem[{\citenamefont{Chatrchyan et~al.}(2013)}]{Chatrchyan:2012jja}
\bibinfo{author}{\bibfnamefont{S.}~\bibnamefont{Chatrchyan}}
  \bibnamefont{et~al.} (\bibinfo{collaboration}{CMS Collaboration}),
  \bibinfo{journal}{Phys.Rev.Lett.} \textbf{\bibinfo{volume}{110}},
  \bibinfo{pages}{081803} (\bibinfo{year}{2013}), \eprint{1212.6639}.

\bibitem[{\citenamefont{Aad et~al.}(2013)}]{Aad:2013xqa}
\bibinfo{author}{\bibfnamefont{G.}~\bibnamefont{Aad}} \bibnamefont{et~al.}
  (\bibinfo{collaboration}{ATLAS Collaboration}), \bibinfo{journal}{Phys.Lett.}
  \textbf{\bibinfo{volume}{B726}}, \bibinfo{pages}{120} (\bibinfo{year}{2013}),
  \eprint{1307.1432}.

\bibitem[{\citenamefont{Berge et~al.}(2008)\citenamefont{Berge, Bernreuther,
  and Ziethe}}]{Berge:2008wi}
\bibinfo{author}{\bibfnamefont{S.}~\bibnamefont{Berge}},
  \bibinfo{author}{\bibfnamefont{W.}~\bibnamefont{Bernreuther}},
  \bibnamefont{and} \bibinfo{author}{\bibfnamefont{J.}~\bibnamefont{Ziethe}},
  \bibinfo{journal}{Phys.Rev.Lett.} \textbf{\bibinfo{volume}{100}},
  \bibinfo{pages}{171605} (\bibinfo{year}{2008}), \eprint{0801.2297}.

\bibitem[{\citenamefont{Berge and Bernreuther}(2009)}]{Berge:2008dr}
\bibinfo{author}{\bibfnamefont{S.}~\bibnamefont{Berge}} \bibnamefont{and}
  \bibinfo{author}{\bibfnamefont{W.}~\bibnamefont{Bernreuther}},
  \bibinfo{journal}{Phys.Lett.} \textbf{\bibinfo{volume}{B671}},
  \bibinfo{pages}{470} (\bibinfo{year}{2009}), \eprint{0812.1910}.

\bibitem[{\citenamefont{Berge et~al.}(2011)\citenamefont{Berge, Bernreuther,
  Niepelt, and Spiesberger}}]{Berge:2011ij}
\bibinfo{author}{\bibfnamefont{S.}~\bibnamefont{Berge}},
  \bibinfo{author}{\bibfnamefont{W.}~\bibnamefont{Bernreuther}},
  \bibinfo{author}{\bibfnamefont{B.}~\bibnamefont{Niepelt}}, \bibnamefont{and}
  \bibinfo{author}{\bibfnamefont{H.}~\bibnamefont{Spiesberger}},
  \bibinfo{journal}{Phys.Rev.} \textbf{\bibinfo{volume}{D84}},
  \bibinfo{pages}{116003} (\bibinfo{year}{2011}), \eprint{1108.0670}.

\bibitem[{\citenamefont{Harnik et~al.}(2013)\citenamefont{Harnik, Martin, Okui,
  Primulando, and Yu}}]{Harnik:2013aja}
\bibinfo{author}{\bibfnamefont{R.}~\bibnamefont{Harnik}},
  \bibinfo{author}{\bibfnamefont{A.}~\bibnamefont{Martin}},
  \bibinfo{author}{\bibfnamefont{T.}~\bibnamefont{Okui}},
  \bibinfo{author}{\bibfnamefont{R.}~\bibnamefont{Primulando}},
  \bibnamefont{and} \bibinfo{author}{\bibfnamefont{F.}~\bibnamefont{Yu}},
  \bibinfo{journal}{Phys.Rev.} \textbf{\bibinfo{volume}{D88}},
  \bibinfo{pages}{076009} (\bibinfo{year}{2013}), \eprint{1308.1094}.

\bibitem[{\citenamefont{Przedzinski et~al.}(2014)\citenamefont{Przedzinski,
  Richter-Was, and Was}}]{Przedzinski:2014pla}
\bibinfo{author}{\bibfnamefont{T.}~\bibnamefont{Przedzinski}},
  \bibinfo{author}{\bibfnamefont{E.}~\bibnamefont{Richter-Was}},
  \bibnamefont{and} \bibinfo{author}{\bibfnamefont{Z.}~\bibnamefont{Was}},
  \bibinfo{journal}{Eur. Phys. J.} \textbf{\bibinfo{volume}{C74}},
  \bibinfo{pages}{3177} (\bibinfo{year}{2014}), \eprint{1406.1647}.

\bibitem[{\citenamefont{Dolan et~al.}(2014)\citenamefont{Dolan, Harris,
  Jankowiak, and Spannowsky}}]{Dolan:2014upa}
\bibinfo{author}{\bibfnamefont{M.~J.} \bibnamefont{Dolan}},
  \bibinfo{author}{\bibfnamefont{P.}~\bibnamefont{Harris}},
  \bibinfo{author}{\bibfnamefont{M.}~\bibnamefont{Jankowiak}},
  \bibnamefont{and}
  \bibinfo{author}{\bibfnamefont{M.}~\bibnamefont{Spannowsky}},
  \bibinfo{journal}{Phys. Rev.} \textbf{\bibinfo{volume}{D90}},
  \bibinfo{pages}{073008} (\bibinfo{year}{2014}), \eprint{1406.3322}.

\bibitem[{\citenamefont{Berge et~al.}(2014)\citenamefont{Berge, Bernreuther,
  and Kirchner}}]{Berge:2014sra}
\bibinfo{author}{\bibfnamefont{S.}~\bibnamefont{Berge}},
  \bibinfo{author}{\bibfnamefont{W.}~\bibnamefont{Bernreuther}},
  \bibnamefont{and} \bibinfo{author}{\bibfnamefont{S.}~\bibnamefont{Kirchner}},
  \bibinfo{journal}{Eur. Phys. J.} \textbf{\bibinfo{volume}{C74}},
  \bibinfo{pages}{3164} (\bibinfo{year}{2014}), \eprint{1408.0798}.

\bibitem[{\citenamefont{Askew et~al.}(2015)\citenamefont{Askew, Jaiswal, Okui,
  Prosper, and Sato}}]{Askew:2015mda}
\bibinfo{author}{\bibfnamefont{A.}~\bibnamefont{Askew}},
  \bibinfo{author}{\bibfnamefont{P.}~\bibnamefont{Jaiswal}},
  \bibinfo{author}{\bibfnamefont{T.}~\bibnamefont{Okui}},
  \bibinfo{author}{\bibfnamefont{H.~B.} \bibnamefont{Prosper}},
  \bibnamefont{and} \bibinfo{author}{\bibfnamefont{N.}~\bibnamefont{Sato}},
  \bibinfo{journal}{Phys. Rev.} \textbf{\bibinfo{volume}{D91}},
  \bibinfo{pages}{075014} (\bibinfo{year}{2015}), \eprint{1501.03156}.

\bibitem[{\citenamefont{Berge et~al.}(2013)\citenamefont{Berge, Bernreuther,
  and Spiesberger}}]{Berge:2013jra}
\bibinfo{author}{\bibfnamefont{S.}~\bibnamefont{Berge}},
  \bibinfo{author}{\bibfnamefont{W.}~\bibnamefont{Bernreuther}},
  \bibnamefont{and}
  \bibinfo{author}{\bibfnamefont{H.}~\bibnamefont{Spiesberger}},
  \bibinfo{journal}{Phys.Lett.} \textbf{\bibinfo{volume}{B727}},
  \bibinfo{pages}{488} (\bibinfo{year}{2013}), \eprint{1308.2674}.

\bibitem[{\citenamefont{Bower et~al.}(2002)\citenamefont{Bower, Pierzchala,
  Was, and Worek}}]{Bower:2002zx}
\bibinfo{author}{\bibfnamefont{G.}~\bibnamefont{Bower}},
  \bibinfo{author}{\bibfnamefont{T.}~\bibnamefont{Pierzchala}},
  \bibinfo{author}{\bibfnamefont{Z.}~\bibnamefont{Was}}, \bibnamefont{and}
  \bibinfo{author}{\bibfnamefont{M.}~\bibnamefont{Worek}},
  \bibinfo{journal}{Phys.Lett.} \textbf{\bibinfo{volume}{B543}},
  \bibinfo{pages}{227} (\bibinfo{year}{2002}), \eprint{hep-ph/0204292}.

\bibitem[{\citenamefont{Desch et~al.}(2003)\citenamefont{Desch, Was, and
  Worek}}]{Desch:2003mw}
\bibinfo{author}{\bibfnamefont{K.}~\bibnamefont{Desch}},
  \bibinfo{author}{\bibfnamefont{Z.}~\bibnamefont{Was}}, \bibnamefont{and}
  \bibinfo{author}{\bibfnamefont{M.}~\bibnamefont{Worek}},
  \bibinfo{journal}{Eur.Phys.J.} \textbf{\bibinfo{volume}{C29}},
  \bibinfo{pages}{491} (\bibinfo{year}{2003}), \eprint{hep-ph/0302046}.

\bibitem[{\citenamefont{Desch et~al.}(2004)\citenamefont{Desch, Imhof, Was, and
  Worek}}]{Desch:2003rw}
\bibinfo{author}{\bibfnamefont{K.}~\bibnamefont{Desch}},
  \bibinfo{author}{\bibfnamefont{A.}~\bibnamefont{Imhof}},
  \bibinfo{author}{\bibfnamefont{Z.}~\bibnamefont{Was}}, \bibnamefont{and}
  \bibinfo{author}{\bibfnamefont{M.}~\bibnamefont{Worek}},
  \bibinfo{journal}{Phys.Lett.} \textbf{\bibinfo{volume}{B579}},
  \bibinfo{pages}{157} (\bibinfo{year}{2004}), \eprint{hep-ph/0307331}.

\bibitem[{\citenamefont{Was and Worek}(2002)}]{Was:2002gv}
\bibinfo{author}{\bibfnamefont{Z.}~\bibnamefont{Was}} \bibnamefont{and}
  \bibinfo{author}{\bibfnamefont{M.}~\bibnamefont{Worek}},
  \bibinfo{journal}{Acta Phys.Polon.} \textbf{\bibinfo{volume}{B33}},
  \bibinfo{pages}{1875} (\bibinfo{year}{2002}), \eprint{hep-ph/0202007}.

\bibitem[{\citenamefont{Worek}(2003)}]{Worek:2003zp}
\bibinfo{author}{\bibfnamefont{M.}~\bibnamefont{Worek}}, \bibinfo{journal}{Acta
  Phys.Polon.} \textbf{\bibinfo{volume}{B34}}, \bibinfo{pages}{4549}
  (\bibinfo{year}{2003}), \eprint{hep-ph/0305082}.

\bibitem[{\citenamefont{Pich and Tuzon}(2009)}]{Pich:2009sp}
\bibinfo{author}{\bibfnamefont{A.}~\bibnamefont{Pich}} \bibnamefont{and}
  \bibinfo{author}{\bibfnamefont{P.}~\bibnamefont{Tuzon}},
  \bibinfo{journal}{Phys. Rev.} \textbf{\bibinfo{volume}{D80}},
  \bibinfo{pages}{091702} (\bibinfo{year}{2009}), \eprint{0908.1554}.

\bibitem[{\citenamefont{Rouge}(1990)}]{Rouge:1990kv}
\bibinfo{author}{\bibfnamefont{A.}~\bibnamefont{Rouge}},
  \bibinfo{journal}{Z.Phys.} \textbf{\bibinfo{volume}{C48}},
  \bibinfo{pages}{75} (\bibinfo{year}{1990}).

\bibitem[{\citenamefont{Davier et~al.}(1993)\citenamefont{Davier, Duflot,
  Le~Diberder, and Rouge}}]{Davier:1992nw}
\bibinfo{author}{\bibfnamefont{M.}~\bibnamefont{Davier}},
  \bibinfo{author}{\bibfnamefont{L.}~\bibnamefont{Duflot}},
  \bibinfo{author}{\bibfnamefont{F.}~\bibnamefont{Le~Diberder}},
  \bibnamefont{and} \bibinfo{author}{\bibfnamefont{A.}~\bibnamefont{Rouge}},
  \bibinfo{journal}{Phys.Lett.} \textbf{\bibinfo{volume}{B306}},
  \bibinfo{pages}{411} (\bibinfo{year}{1993}).

\bibitem[{\citenamefont{Kuhn}(1995)}]{Kuhn:1995nn}
\bibinfo{author}{\bibfnamefont{J.~H.} \bibnamefont{Kuhn}},
  \bibinfo{journal}{Phys.Rev.} \textbf{\bibinfo{volume}{D52}},
  \bibinfo{pages}{3128} (\bibinfo{year}{1995}), \eprint{hep-ph/9505303}.

\bibitem[{\citenamefont{Stahl}(2000)}]{Stahl:2000aq}
\bibinfo{author}{\bibfnamefont{A.}~\bibnamefont{Stahl}},
  \bibinfo{journal}{Springer Tracts Mod.Phys.} \textbf{\bibinfo{volume}{160}},
  \bibinfo{pages}{1} (\bibinfo{year}{2000}).

\bibitem[{\citenamefont{van Hameren}(2009)}]{vanHameren:2007pt}
\bibinfo{author}{\bibfnamefont{A.}~\bibnamefont{van Hameren}},
  \bibinfo{journal}{Acta Phys. Polon.} \textbf{\bibinfo{volume}{B40}},
  \bibinfo{pages}{259} (\bibinfo{year}{2009}), \eprint{0710.2448}.

\bibitem[{\citenamefont{Vermaseren}(2000)}]{Vermaseren:2000nd}
\bibinfo{author}{\bibfnamefont{J.~A.~M.} \bibnamefont{Vermaseren}}
  (\bibinfo{year}{2000}), \eprint{math-ph/0010025}.

\bibitem[{\citenamefont{Buckley et~al.}(2015)\citenamefont{Buckley, Ferrando,
  Lloyd, Nordstroem, Page, Ruefenacht, Schoenherr, and Watt}}]{Buckley:2014ana}
\bibinfo{author}{\bibfnamefont{A.}~\bibnamefont{Buckley}},
  \bibinfo{author}{\bibfnamefont{J.}~\bibnamefont{Ferrando}},
  \bibinfo{author}{\bibfnamefont{S.}~\bibnamefont{Lloyd}},
  \bibinfo{author}{\bibfnamefont{K.}~\bibnamefont{Nordstroem}},
  \bibinfo{author}{\bibfnamefont{B.}~\bibnamefont{Page}},
  \bibinfo{author}{\bibfnamefont{M.}~\bibnamefont{Ruefenacht}},
  \bibinfo{author}{\bibfnamefont{M.}~\bibnamefont{Schoenherr}},
  \bibnamefont{and} \bibinfo{author}{\bibfnamefont{G.}~\bibnamefont{Watt}},
  \bibinfo{journal}{Eur. Phys. J.} \textbf{\bibinfo{volume}{C75}},
  \bibinfo{pages}{132} (\bibinfo{year}{2015}), \eprint{1412.7420}.

\bibitem[{\citenamefont{Gough}(2009)}]{Gough:2009:GSL:1538674}
\bibinfo{author}{\bibfnamefont{B.}~\bibnamefont{Gough}},
  \emph{\bibinfo{title}{GNU Scientific Library Reference Manual - Third
  Edition}} (\bibinfo{publisher}{Network Theory Ltd.}, \bibinfo{year}{2009}),
  \bibinfo{edition}{3rd} ed., ISBN \bibinfo{isbn}{0954612078, 9780954612078}.

\bibitem[{\citenamefont{Brun and Rademakers}(1997)}]{Brun:1997pa}
\bibinfo{author}{\bibfnamefont{R.}~\bibnamefont{Brun}} \bibnamefont{and}
  \bibinfo{author}{\bibfnamefont{F.}~\bibnamefont{Rademakers}},
  \bibinfo{journal}{Nucl. Instrum. Meth.} \textbf{\bibinfo{volume}{A389}},
  \bibinfo{pages}{81} (\bibinfo{year}{1997}).

\bibitem[{\citenamefont{Campbell and Ellis}(2010)}]{Campbell:2010ff}
\bibinfo{author}{\bibfnamefont{J.~M.} \bibnamefont{Campbell}} \bibnamefont{and}
  \bibinfo{author}{\bibfnamefont{R.~K.} \bibnamefont{Ellis}},
  \bibinfo{journal}{Nucl. Phys. Proc. Suppl.}
  \textbf{\bibinfo{volume}{205-206}}, \bibinfo{pages}{10}
  (\bibinfo{year}{2010}), \eprint{1007.3492}.

\bibitem[{\citenamefont{ATLAS-collaboration}(2008)}]{ATLAS-S08003}
\bibinfo{author}{\bibnamefont{ATLAS-collaboration}}, \bibinfo{journal}{Journal
  of Instrumentation} \textbf{\bibinfo{volume}{3}}, \bibinfo{pages}{S08003}
  (\bibinfo{year}{2008}).

\bibitem[{\citenamefont{ATLAS-collaboration}(2013)}]{ATLtauconf}
\bibinfo{author}{\bibnamefont{ATLAS-collaboration}},
  \bibinfo{journal}{ATLAS-CONF-2013-108, ATLAS-COM-CONF-2013-095}
  (\bibinfo{year}{2013}).

\bibitem[{\citenamefont{Arbey et~al.}(2015)\citenamefont{Arbey, Ellis, Godbole,
  and Mahmoudi}}]{Arbey:2014msa}
\bibinfo{author}{\bibfnamefont{A.}~\bibnamefont{Arbey}},
  \bibinfo{author}{\bibfnamefont{J.}~\bibnamefont{Ellis}},
  \bibinfo{author}{\bibfnamefont{R.~M.} \bibnamefont{Godbole}},
  \bibnamefont{and} \bibinfo{author}{\bibfnamefont{F.}~\bibnamefont{Mahmoudi}},
  \bibinfo{journal}{Eur. Phys. J.} \textbf{\bibinfo{volume}{C75}},
  \bibinfo{pages}{85} (\bibinfo{year}{2015}), \eprint{1410.4824}.

\bibitem[{\citenamefont{Li and Wagner}(2015)}]{Li:2015yla}
\bibinfo{author}{\bibfnamefont{B.}~\bibnamefont{Li}} \bibnamefont{and}
  \bibinfo{author}{\bibfnamefont{C.~E.~M.} \bibnamefont{Wagner}},
  \bibinfo{journal}{Phys. Rev.} \textbf{\bibinfo{volume}{D91}},
  \bibinfo{pages}{095019} (\bibinfo{year}{2015}), \eprint{1502.02210}.

\bibitem[{\citenamefont{King et~al.}(2015)\citenamefont{King, Muhlleitner,
  Nevzorov, and Walz}}]{King:2015oxa}
\bibinfo{author}{\bibfnamefont{S.~F.} \bibnamefont{King}},
  \bibinfo{author}{\bibfnamefont{M.}~\bibnamefont{Muhlleitner}},
  \bibinfo{author}{\bibfnamefont{R.}~\bibnamefont{Nevzorov}}, \bibnamefont{and}
  \bibinfo{author}{\bibfnamefont{K.}~\bibnamefont{Walz}}
  (\bibinfo{year}{2015}), \eprint{1508.03255}.

\bibitem[{\citenamefont{Branco et~al.}(2012)\citenamefont{Branco, Ferreira,
  Lavoura, Rebelo, Sher, and Silva}}]{Branco:2011iw}
\bibinfo{author}{\bibfnamefont{G.~C.} \bibnamefont{Branco}},
  \bibinfo{author}{\bibfnamefont{P.~M.} \bibnamefont{Ferreira}},
  \bibinfo{author}{\bibfnamefont{L.}~\bibnamefont{Lavoura}},
  \bibinfo{author}{\bibfnamefont{M.~N.} \bibnamefont{Rebelo}},
  \bibinfo{author}{\bibfnamefont{M.}~\bibnamefont{Sher}}, \bibnamefont{and}
  \bibinfo{author}{\bibfnamefont{J.~P.} \bibnamefont{Silva}},
  \bibinfo{journal}{Phys. Rept.} \textbf{\bibinfo{volume}{516}},
  \bibinfo{pages}{1} (\bibinfo{year}{2012}), \eprint{1106.0034}.

\bibitem[{\citenamefont{Dawson et~al.}(2013)}]{Dawson:2013bba}
\bibinfo{author}{\bibfnamefont{S.}~\bibnamefont{Dawson}} \bibnamefont{et~al.},
  in \emph{\bibinfo{booktitle}{{Community Summer Study 2013: Snowmass on the
  Mississippi (CSS2013) Minneapolis, MN, USA, July 29-August 6, 2013}}}
  (\bibinfo{year}{2013}), \eprint{1310.8361}.

\bibitem[{\citenamefont{Jung and Pich}(2014)}]{Jung:2013hka}
\bibinfo{author}{\bibfnamefont{M.}~\bibnamefont{Jung}} \bibnamefont{and}
  \bibinfo{author}{\bibfnamefont{A.}~\bibnamefont{Pich}},
  \bibinfo{journal}{JHEP} \textbf{\bibinfo{volume}{04}}, \bibinfo{pages}{076}
  (\bibinfo{year}{2014}), \eprint{1308.6283}.

\bibitem[{\citenamefont{Celis et~al.}(2013)\citenamefont{Celis, Ilisie, and
  Pich}}]{Celis:2013ixa}
\bibinfo{author}{\bibfnamefont{A.}~\bibnamefont{Celis}},
  \bibinfo{author}{\bibfnamefont{V.}~\bibnamefont{Ilisie}}, \bibnamefont{and}
  \bibinfo{author}{\bibfnamefont{A.}~\bibnamefont{Pich}},
  \bibinfo{journal}{JHEP} \textbf{\bibinfo{volume}{12}}, \bibinfo{pages}{095}
  (\bibinfo{year}{2013}), \eprint{1310.7941}.

\bibitem[{\citenamefont{Bernreuther et~al.}(1992)\citenamefont{Bernreuther,
  Schroder, and Pham}}]{Bernreuther:1992dz}
\bibinfo{author}{\bibfnamefont{W.}~\bibnamefont{Bernreuther}},
  \bibinfo{author}{\bibfnamefont{T.}~\bibnamefont{Schroder}}, \bibnamefont{and}
  \bibinfo{author}{\bibfnamefont{T.}~\bibnamefont{Pham}},
  \bibinfo{journal}{Phys.Lett.} \textbf{\bibinfo{volume}{B279}},
  \bibinfo{pages}{389} (\bibinfo{year}{1992}).

\bibitem[{\citenamefont{Accomando et~al.}(2006)}]{Accomando:2006ga}
\bibinfo{author}{\bibfnamefont{E.}~\bibnamefont{Accomando}}
  \bibnamefont{et~al.} (\bibinfo{year}{2006}), \eprint{hep-ph/0608079}.

\bibitem[{\citenamefont{Eberhardt et~al.}(2013)\citenamefont{Eberhardt,
  Nierste, and Wiebusch}}]{Eberhardt:2013uba}
\bibinfo{author}{\bibfnamefont{O.}~\bibnamefont{Eberhardt}},
  \bibinfo{author}{\bibfnamefont{U.}~\bibnamefont{Nierste}}, \bibnamefont{and}
  \bibinfo{author}{\bibfnamefont{M.}~\bibnamefont{Wiebusch}},
  \bibinfo{journal}{JHEP} \textbf{\bibinfo{volume}{07}}, \bibinfo{pages}{118}
  (\bibinfo{year}{2013}), \eprint{1305.1649}.

\bibitem[{\citenamefont{Barroso et~al.}(2012)\citenamefont{Barroso, Ferreira,
  Santos, and Silva}}]{Barroso:2012wz}
\bibinfo{author}{\bibfnamefont{A.}~\bibnamefont{Barroso}},
  \bibinfo{author}{\bibfnamefont{P.~M.} \bibnamefont{Ferreira}},
  \bibinfo{author}{\bibfnamefont{R.}~\bibnamefont{Santos}}, \bibnamefont{and}
  \bibinfo{author}{\bibfnamefont{J.~P.} \bibnamefont{Silva}},
  \bibinfo{journal}{Phys. Rev.} \textbf{\bibinfo{volume}{D86}},
  \bibinfo{pages}{015022} (\bibinfo{year}{2012}), \eprint{1205.4247}.

\bibitem[{\citenamefont{Cheung et~al.}(2014)\citenamefont{Cheung, Lee, Senaha,
  and Tseng}}]{Cheung:2014oaa}
\bibinfo{author}{\bibfnamefont{K.}~\bibnamefont{Cheung}},
  \bibinfo{author}{\bibfnamefont{J.~S.} \bibnamefont{Lee}},
  \bibinfo{author}{\bibfnamefont{E.}~\bibnamefont{Senaha}}, \bibnamefont{and}
  \bibinfo{author}{\bibfnamefont{P.-Y.} \bibnamefont{Tseng}},
  \bibinfo{journal}{JHEP} \textbf{\bibinfo{volume}{06}}, \bibinfo{pages}{149}
  (\bibinfo{year}{2014}), \eprint{1403.4775}.

\bibitem[{\citenamefont{Grzadkowski et~al.}(2014)\citenamefont{Grzadkowski,
  Ogreid, and Osland}}]{Grzadkowski:2014ada}
\bibinfo{author}{\bibfnamefont{B.}~\bibnamefont{Grzadkowski}},
  \bibinfo{author}{\bibfnamefont{O.~M.} \bibnamefont{Ogreid}},
  \bibnamefont{and} \bibinfo{author}{\bibfnamefont{P.}~\bibnamefont{Osland}},
  \bibinfo{journal}{JHEP} \textbf{\bibinfo{volume}{11}}, \bibinfo{pages}{084}
  (\bibinfo{year}{2014}), \eprint{1409.7265}.

\bibitem[{\citenamefont{Fontes et~al.}(2015{\natexlab{a}})\citenamefont{Fontes,
  Romao, Santos, and Silva}}]{Fontes:2015mea}
\bibinfo{author}{\bibfnamefont{D.}~\bibnamefont{Fontes}},
  \bibinfo{author}{\bibfnamefont{J.~C.} \bibnamefont{Romao}},
  \bibinfo{author}{\bibfnamefont{R.}~\bibnamefont{Santos}}, \bibnamefont{and}
  \bibinfo{author}{\bibfnamefont{J.~P.} \bibnamefont{Silva}},
  \bibinfo{journal}{JHEP} \textbf{\bibinfo{volume}{06}}, \bibinfo{pages}{060}
  (\bibinfo{year}{2015}{\natexlab{a}}), \eprint{1502.01720}.

\bibitem[{\citenamefont{Inoue et~al.}(2014)\citenamefont{Inoue, Ramsey-Musolf,
  and Zhang}}]{Inoue:2014nva}
\bibinfo{author}{\bibfnamefont{S.}~\bibnamefont{Inoue}},
  \bibinfo{author}{\bibfnamefont{M.~J.} \bibnamefont{Ramsey-Musolf}},
  \bibnamefont{and} \bibinfo{author}{\bibfnamefont{Y.}~\bibnamefont{Zhang}},
  \bibinfo{journal}{Phys. Rev.} \textbf{\bibinfo{volume}{D89}},
  \bibinfo{pages}{115023} (\bibinfo{year}{2014}), \eprint{1403.4257}.

\bibitem[{\citenamefont{Fontes et~al.}(2015{\natexlab{b}})\citenamefont{Fontes,
  Romao, Santos, and Silva}}]{Fontes:2015xva}
\bibinfo{author}{\bibfnamefont{D.}~\bibnamefont{Fontes}},
  \bibinfo{author}{\bibfnamefont{J.~C.} \bibnamefont{Romao}},
  \bibinfo{author}{\bibfnamefont{R.}~\bibnamefont{Santos}}, \bibnamefont{and}
  \bibinfo{author}{\bibfnamefont{J.~P.} \bibnamefont{Silva}}
  (\bibinfo{year}{2015}{\natexlab{b}}), \eprint{1506.06755}.

\bibitem[{\citenamefont{Chen et~al.}(2015)\citenamefont{Chen, Dawson, and
  Zhang}}]{Chen:2015gaa}
\bibinfo{author}{\bibfnamefont{C.-Y.} \bibnamefont{Chen}},
  \bibinfo{author}{\bibfnamefont{S.}~\bibnamefont{Dawson}}, \bibnamefont{and}
  \bibinfo{author}{\bibfnamefont{Y.}~\bibnamefont{Zhang}},
  \bibinfo{journal}{JHEP} \textbf{\bibinfo{volume}{06}}, \bibinfo{pages}{056}
  (\bibinfo{year}{2015}), \eprint{1503.01114}.

\bibitem[{\citenamefont{Baker et~al.}(2006)}]{Baker:2006ts}
\bibinfo{author}{\bibfnamefont{C.~A.} \bibnamefont{Baker}}
  \bibnamefont{et~al.}, \bibinfo{journal}{Phys. Rev. Lett.}
  \textbf{\bibinfo{volume}{97}}, \bibinfo{pages}{131801}
  (\bibinfo{year}{2006}), \eprint{hep-ex/0602020}.

\bibitem[{\citenamefont{Baron et~al.}(2014)}]{Baron:2013eja}
\bibinfo{author}{\bibfnamefont{J.}~\bibnamefont{Baron}} \bibnamefont{et~al.}
  (\bibinfo{collaboration}{ACME}), \bibinfo{journal}{Science}
  \textbf{\bibinfo{volume}{343}}, \bibinfo{pages}{269} (\bibinfo{year}{2014}),
  \eprint{1310.7534}.

\bibitem[{\citenamefont{Aoki et~al.}(2009)\citenamefont{Aoki, Kanemura,
  Tsumura, and Yagyu}}]{Aoki:2009ha}
\bibinfo{author}{\bibfnamefont{M.}~\bibnamefont{Aoki}},
  \bibinfo{author}{\bibfnamefont{S.}~\bibnamefont{Kanemura}},
  \bibinfo{author}{\bibfnamefont{K.}~\bibnamefont{Tsumura}}, \bibnamefont{and}
  \bibinfo{author}{\bibfnamefont{K.}~\bibnamefont{Yagyu}},
  \bibinfo{journal}{Phys. Rev.} \textbf{\bibinfo{volume}{D80}},
  \bibinfo{pages}{015017} (\bibinfo{year}{2009}), \eprint{0902.4665}.

\end{thebibliography}

\end{document}